\documentclass[12pt,numbers]{elsarticle}
\usepackage[utf8]{inputenc}
\usepackage[english]{babel}
\usepackage{amsmath}
\usepackage{amsfonts}
\usepackage{amssymb}
\usepackage{amsopn}
\usepackage{amsthm}
\usepackage{bm}
\usepackage{makeidx}
\usepackage{graphicx}
\usepackage{grffile}
\usepackage{dsfont}
\usepackage{algorithmic}
\journal{arXiv}

\newtheorem*{mythm}{Theorem}

\begin{document}

\begin{frontmatter}

\title{Geometric Perturbation Theory and Acoustic Boundary Condition Dynamics}
\author[auth]{David T. Heider}
\ead{david.heider@tum.de}
\author[auth]{J. Leo van Hemmen\corref{cor}}
\ead{lvh@tum.de}
\cortext[cor]{Corresponding author}
\address[auth]{Physik Department T35, Technische Universit\"{a}t M\"{u}nchen, \\ 85747 Garching bei M\"{u}nchen, Germany}

\begin{abstract}
Geometric perturbation theory is universally needed but not recognized as such yet. A typical example is provided by the three-dimensional wave equation, widely used in acoustics. We face vibrating eardrums as binaural auditory input and stemming from an external sound source. In the setup of internally coupled ears (ICE), which are present in more than half of the land-living vertebrates, the two tympana are coupled by an internal air-filled cavity, whose geometry determines the acoustic properties of the ICE system. The eardrums themselves are described by a two-dimensional, damped, wave equation and are part of the spatial boundary conditions of the three-dimensional Laplacian belonging to the wave equation in the internal cavity that couples and internally drives the eardrums. In animals with ICE the resulting signal is the superposition of external sound arriving at both eardrums and the internal pressure coupling them. This is also the typical setup for geometric perturbation theory. In the context of ICE it boils down to acoustic boundary-condition dynamics (ABCD) for the coupled dynamical system of eardrums and internal cavity. In acoustics the deviations from equilibrium are extremely small (nm range). Perturbation theory therefore seems natural and is shown to be appropriate. In doing so, we use a time-dependent perturbation theory \`a la Dirac in the context of Duhamel's principle. The relaxation dynamics of the tympanic-membrane system, which neuronal information processing stems from, is explicitly obtained in first order. Furthermore, both the initial and the quasi-stationary asymptotic state are derived and analyzed. Introducing the so-called spinning parameter as a natural generalization of the modal cut-off criterion in duct acoustics, we justify the piston approximation. Finally, we set the general stage for geometric perturbation theory where $(d-1)$-dimensional manifolds as subsets of the boundary of a $d$-dimensional domain are driven by their own dynamics with the domain pressure $p$ and an external source term as input, at the same time constituting time-dependent boundary conditions for $p$. 
\end{abstract}


\end{frontmatter}

\section{Introduction: Geometric perturbation theory}
\label{intro}

Though omnipresent, geometric perturbation theory has hardly been discerned as such, if at all. We start by describing a typical situation, viz., that of internally coupled ears (ICE), which is present in more than half of the terrestrial vertebrates. Though ICE has been observed and described since long, it has attracted considerable attention only recently \cite{ice-editorial}.  It is used here as a typical example from acoustics to illustrate the general idea of geometric perturbation theory. 

Terrestrial animals perform azimuthal sound localization through neuronally determining the time difference between left and right eardrums, the so-called interaural time difference (ITD). Whereas mammals have independent ears, which do not influence each other, most of the terrestrial vertebrates have ICE at their disposal, which means an internal, air-filled, cavity that connects the two eardrums. ICE allows in particular small animals with small head and, hence, small ITD to greatly increase its effective interaural distance; typically, for low frequencies by a multiplicative factor of 2--4 \cite{ice-editorial}. What the animal then actually perceives is the so-called the \emph{internal} time difference (iTD) as the superposition of the external auditory stimulus $p_{\rm ext}$ operating at the two eardrums located at, say, $x=0$ and $x=L$, and the internal pressure $p$; see Fig.~\ref{iTD}.  

The eardrums are driven by the superposition of  the external stimulus $p_{\rm ext}$ and the internal pressure $p$ that so-to-speak couples them, and are part of the the boundary confining the wave equation governing the internal pressure $p$.  In the, for the sake of clarity, concrete situation of ICE analyzed here in Fig.~\ref{iTD}\,(e) as well as in any other acoustic problem, the time-dependent deviations from equilibrium (defined to be $u(\mathbf{x}) =0$ for all $\mathbf{x}$ on -- sloppily -- the eardrums; see below) are extremely small, in the nm range and thus orders of magnitude smaller than any other physical quantity involved. Furthermore and contrary to naive physical intuition, eardrums are strongly damped. Since an exact solution to the coupled dynamics of outside stimulus, inside pressure, and both eardrums is nearly always out of range and the tympanic deviations from equilibrium are small (nm), a perturbation theory seems natural. 

As early as 1938, Herbert Fr\"{o}hlich \cite{Frohlich} published a short paper with the title ``A solution of the Schr\"{o}dinger equation by a perturbation of the boundary conditions," where he initiated a perturbation theory for the change in eigenfunctions and eigenvalues by changing the boundary conditions but still fixing the domain, and also by changing the domain as well. Both types of change, however, did not depend on time. In the acoustic context of the above examples, they do and we are also interested in a general treatment of the ensuing time evolution but, despite the small amplitudes involved, until now a systematic, time-dependent, perturbation theory did not exist. That is what we do here. 

Starting with the work of Beale and Rosencrans \cite{Beale1}, which Beale \cite{Beale2} has extended and worked out, the approach was different. In view of the nm deviations from equilibrium, the boundary was fixed to be the original one and assumed to be ``locally reacting'' in the sense that is was covered by independent oscillators satisfying a separate damped harmonic-oscillator equation coupled linearly to $p$ \cite{Beale2,Beale1}. Beale and Rosencrans \cite{Beale1} called the final construct Acoustic Boundary Conditions or, for short, ABC. In so doing they have started the mathematical analysis of ABC. Though their construct can be handled mathematically and Beale and Rosencrans \cite{Beale1} and particularly Beale \cite{Beale2} could prove the existence of a dynamical evolution, it does not correspond to the underlying physics. Nor was it possible to obtain any explicit solution, valuable as it is for practical work. In fact, many concrete situations in vibro-acoustics have been analyzed in a similar spirit \cite{DengLi,semiop2,Pan2, Pan1} -- to name just a few outstanding papers, though even these allow neither a dynamics nor a dynamical coupling either. 

Here we return to the essential physics of the problem by taking the volume fluctuations due to the dynamics of the eardrums as our starting point and incorporating them as time-dependent perturbations of the Laplacian, a procedure that seems to be novel and, more importantly, allows for a systematic perturbation theory. We develop a time-dependent perturbation theory in the style of Dirac \cite{Dirac1,Dirac2}, whose key idea is nicely described by Dirac himself in his classic on quantum mechanics \cite[\S{44}]{Dirac2}. Not only do we present a mathematical perturbation theory for handling time-varying domains and allowing for explicit solutions but also for obtaining the full dynamical evolution for all times $t \ge 0$, including the asymptotics as $t \to \infty$. In a sense, we extend Beale's ABC \cite{Beale2,Beale1} to ABCD, i.e., Acoustic Boundary Condition Dynamics, and in so doing proceed to geometric perturbation theory as the proper generalization. In the context of geometric perturbation theory, a few of ABCD's applications are presented in sections \ref{appl} and \ref{applac}. A summary in terms of a concrete theorem as well as an algorithmic delineation can be found at the end of this paper in section \ref{discas}. 

Before embarking on the mathematical theory, a cautionary remark is in order. The tenor of the present paper is somewhat unconventional in that its goal is two-fold. First, we replace the traditional, i.e., weak \cite{Evans,sh+l}, form of the present fluctuating-boundary value problem, which is rather clumsy (understatement), by a strong form that allows for a systematic and slick perturbation  expansion taking care of the fluctuating boundary with its inherent internal dynamics and exploiting the smallness of its amplitude, which is typical to a huge class of problems in acoustics. In so doing we will also derive a systematic approximation to the exact dynamics, which until now was out of range as well. Second, though Dirac's time-dependent perturbation theory is well known, we will make a small detour to explain a Duhamel version that in the present context is more efficient but less known.

\subsection{ICE and Acoustic Boundary-Condition Dynamics}
\label{ICE+ABCD}

\begin{figure}
\centering
\includegraphics[width=.99\linewidth]{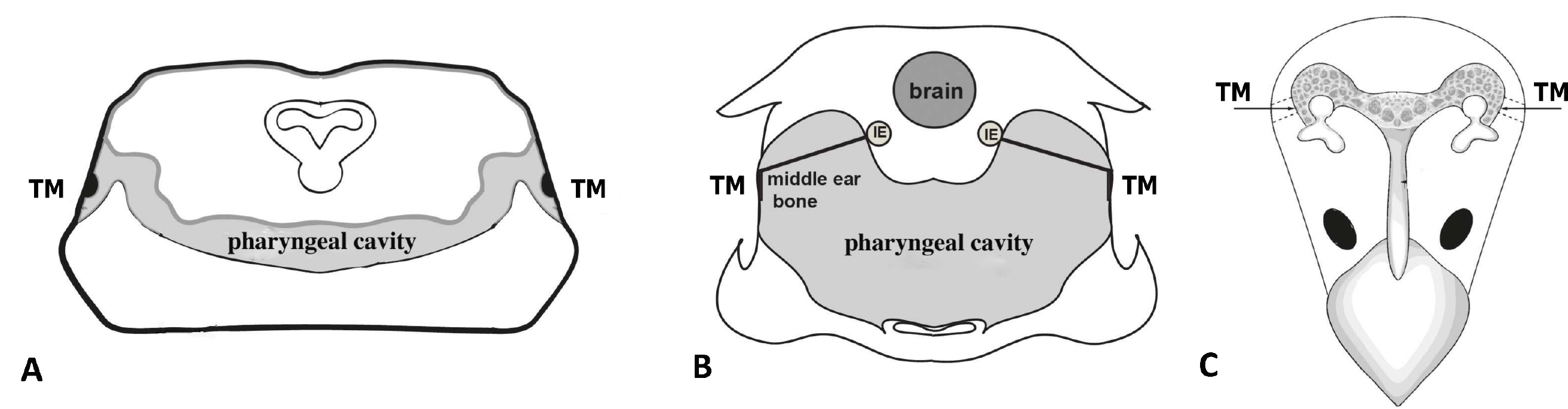}
\includegraphics[width=0.30\linewidth]{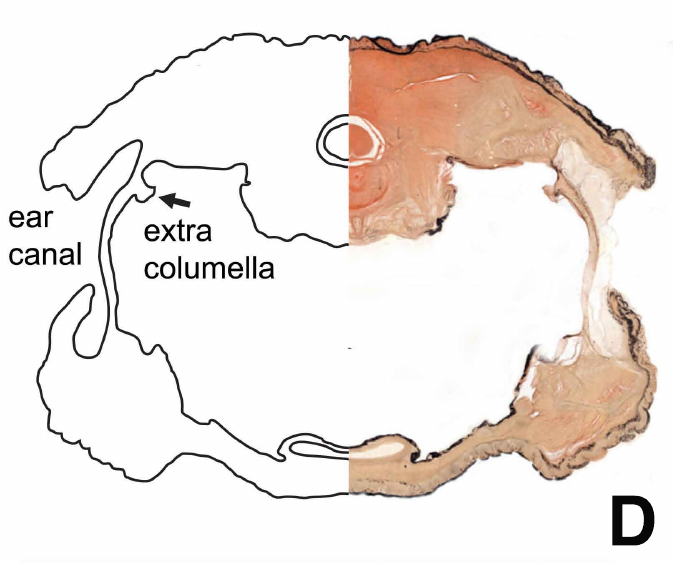}
\qquad
\includegraphics[width = 0.4\textwidth]{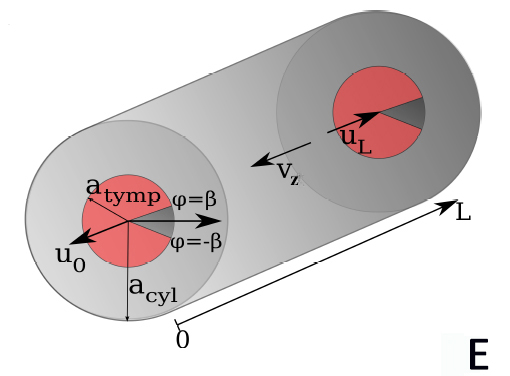}
\caption{Schematic representation of Internally Coupled Ears (ICE) in frogs (a), lizards (b), and birds (c). The bird in (c) is seen from the top, the other two in (a) and (b) show a cross section, and all three exhibit the interaural cavity as a gray tube. (d) Cross section of a real \emph{Gecko} cavity as it occurs in Nature. The extracolumella in (d) is embedded in the eardrum and the beginning of the middle-ear bone in (b) that picks up the eardrum vibrations and transports them through the ear-filled cavity to the cochlea. The latter is quite near to the neuronal information processing of sound in the brain.  The mathematical idealization of (d) is depicted in (e) where the cylindrical cavity $\Omega$ of length $L$ and radius $a_{\rm{cyl}}$ are as used in the ICE model \cite{ICE2} that is used here as example. The circular sectors $\Gamma_0$ and $\Gamma_L$ with radius $a_{\rm{tymp}} \le a_{\rm{cyl}}$ and opening angle $\beta$ are situated at $x=0$ and $x=L$. The left and right membrane-displacement fields are indicated by $u_0$ and $u_L$, respectively. The arrow at the bottom of Fig.~\ref{iTD}\,(e) denotes orientation in positive $x$-direction. Figures \ref{iTD}\,(a)--(c) have been adapted from Christensen-Dalsgaard \cite{jcd1}, (d) stems from Carr, Tang, and Christensen-Dalsgaard \cite{cecytjcd}, and (e) has been taken from Vedurmudi et al. \cite{ICE2}.}
\label{iTD}
\end{figure}

The ICE model \cite{ICE2, ICE3, ICE1} describing internally coupled ears is a geometrical model of binaural directional hearing in lizards as representatives of more than half of the terrestrial vertebrates. Physically, the lizard's eardrums, also termed tympana, are connected by a cylindrical cavity. We model the eardrum cavity by a cylinder of length  $L$ and radius $a_{\rm{cyl}}$. The two sectorial membranes with radius $a_{\rm{tymp}}<a_{\rm{cyl}}$ and sectorial opening angle $2\beta$ are situated at the $x=0/L$ end of the cylindrical cavity; see Fig.~\ref{iTD}. Mathematically, the membranes are denoted by their respective displacement fields $u_{0/L}$ from their equilibrium positions $\Gamma_{0/L}\times\lbrace x=0/L\rbrace=[0,a_{\rm{tymp}}<a_{\rm{cyl}}]\times[\beta;2\pi-\beta]\times\lbrace x=0/L\rbrace$; cf. Fig.~\ref{iTD}E. 
The eardrums respond to the difference between the outside pressure $p_{\rm{ex}}$ and the inside pressure $p$. They detect the difference of an incident plane-wave acoustic signal $p_{\rm{ex}}=p_0 \exp[{i(\omega t-k\Delta)}]$ where $p_{\rm{ex}}$ is the external pressure at $x=0$ or $x=L$, and the acoustic pressure $p$ inside the cylindrical, interaural cavity $\Omega(t)$. Our geometric setup to be used throughout what follows is shown in Fig.~\ref{iTD}\,(e).


We take $u_0:\Gamma_0\to\mathbb{C}$ and $u_L:\Gamma_L\to\mathbb{C}$ to denote the complex-valued membrane displacements at the $x=0$ and $x=L$ end caps of the stationary cylinder $\Omega_0$. See Fig.~\ref{iTD}\,(e), where the tympanum is restricted to the circle with radius $a_\mathrm{tymp} < a_\mathrm{cyl}$. At rest the eardrums are described by $\Gamma_0 = \{x=0, 0 \le r = \sqrt{y^2 + z^2} \le a_\mathrm{tymp} \}$ and $\Gamma_L= \{x=L, 0 \le r = \sqrt{y^2 + z^2} \le a_\mathrm{tymp} \}$, respectively. During a sound stimulus the time-dependent \emph{deviations} along the $x$-axis (no sound, or rest, means deviation $=0$) are denoted by $u_0(t,y,z)$ and $u_L(t,y,z)$ for the eardrums contained in the discs at $x=0$ and $x=L$, respectively, which we have just specified; or simply glance Fig.~\ref{iTD}\,(e). As usual, $t$ stands for time. Since under natural circumstances disco-sound stimuli are exceptional, we can safely assume that eardrum deviations from rest are in the nm range. That is to say, ``small" as compared to any other macroscopic object such as $a_\mathrm{tymp}$ or $a_\mathrm{cyl}$ or $L$.

The time-varying cylindrical cavity is now defined as follows: $\Omega(t)\equiv\lbrace (x,y,z) \in \mathbb{D}^2_{a_{\rm cyl}} \times \mathbb{R} \vert u_0(t,y,z) \leq x = x(y,z) \leq L + u_L(t,y,z)\rbrace$. The cylindrical shapes of the cavity $\Omega_0$ and of the equilibrium positions $\Gamma_{0}$ and $\Gamma_L$ suggest using polar coordinates in the $(y,z)$-plane instead of Cartesian coordinates. Since the treatment of $u_0$ and $u_L$ will proceed completely analogously for both, we use the notation $u_{0/L}=u_k$ where $k\in\lbrace 0,L\rbrace$, mimicking a subscript notation with subscript $\pm$.

\paragraph{Qualitative view on auditory ICE dynamics} As a response to the pressure difference, the membranes start vibrating. The acoustic wave equation for $p$ contains a three-dimensional Laplacian. Because of membrane vibrations, the Laplacian is defined on  a time-varying domain $\Omega(t)$. We indicate this time dependence in the notation by denoting the Laplacian on $\Omega(t)$ by $\Delta_t$ and the Laplacian on $\Omega_0=\Omega(t=0)$ by a single $\Delta$. On $\Omega_0$, the pressure is described by the acoustic wave equation,
\begin{align}
\dfrac{\partial^2 p}{\partial t^2}-c^2\Delta p = 0,
\end{align}
in conjunction with Neumann boundary conditions \cite[chapter 2]{Temkin}; namely, $\partial_{\mathbf{n}}p=0$ where $\mathbf{n}$ denotes the outward unit normal to $\partial\Omega_0$.

The investigation of the model at times $t>0$ requires the three-dimensional Laplacian $\Delta_t$ to live in a time-varying domain based on $\Omega(t) \subset \mathbb{R}^3$. Misusing our language a bit, $\Delta_2$ is simply two-dimensional. Further domain questions will be discussed at the beginning of section~\ref{patag}. 
The full ICE equations \cite{ICE2, ICE3} read in the present notation
\begin{align}
\dfrac{\partial^2 p}{\partial t^2}-c^2\Delta_t p &=0 \ , \label{3DW} \\
\dfrac{\partial^2 u_{0/L}}{\partial t^2}+2\alpha\dfrac{\partial u_{0/L}}{\partial t}-c^2_m\Delta_2 u_{0/L}&=\left.\dfrac{p_{\rm{ex}}-p}{\rho_m d}\right\vert_{x=u(t,\Gamma_{0/L})} 
\label{2DW}.
\end{align}
Equation (\ref{3DW}) describes the wave equation for the cavity pressure $p$ whereas the equations for $u_{0/L}$ in (\ref{2DW}) characterize the vibration pattern of the left ($x=0$) and right ($x=L$) tympanic membranes in response to the difference of the outside pressure signals $p_{\rm{ex}}$ due to, say, a sound signal and the internal cavity pressure $p$; cf. Fig.~\ref{iTD}.
The boundary conditions corresponding to the above equations of motion (\ref{3DW}) and (\ref{2DW}) will be expounded in and after Eq.~(\ref{boundcon}).

The factor $1/\rho_m d$ in (\ref{2DW}) contains two material constants, the membrane density $\rho_m$ and its thickness $d$. Though it is tempting to choose units in such a way that $\rho_m d = 1$, we will see -- cf. Eq.~(\ref{cpl}) -- that the fraction of the density $\rho_0$ of air and $\rho_m$ of the tympanic membrane, which couples outside air to inside air and is such that $\rho_0/\rho_m \approx 10^{-3} \ll 1$, is the relevant coupling parameter. Accordingly, we refrain from putting $\rho_m d = 1$. Damping is incorporated through a phenomenological damping coefficient $\alpha >0$, with dimension s$^{-1}$. Since long it describes reality surprisingly well but has never been derived (yet). In cochlear acoustics its decay time constant is amazingly short; in general, $< 5$~ms. 

By the manifest time dependence of the Laplacian $\Delta_t$ because of the time dependence of its domain $\Omega(t)$ \emph{and} its boundary conditions as exhibited in (\ref{boundcon}), the acoustic wave equation cannot be separated exactly. As shown below, we need to analyze the time-varying boundary conditions with great care. We also note that for $\alpha > 0$ and without external pressure, i.e., $p_{\rm{ex}}\equiv 0$, the ICE equations (\ref{3DW}) \& (\ref{2DW}) approach  the asymptotic solutions $p=0$ and $u_{0/L}=0$ as $t \to \infty$, where $u_{0/L}$ denotes $u_0=0$ or $u_L=0$. This dissipative behavior has been demonstrated with the aid of a Lyapunov functional \cite{Beale2, Beale1}.


The models proposed by Beale and Rosencrans \cite{Beale2, Beale1} differ from (\ref{3DW}) \& (\ref{2DW}) in the sense that local interactions between the constituents of the membranes are not considered. In physical words, the boundary equations describe each point on the undulating surface as a (damped) harmonic oscillator that has no interactions with its neighbors and operates in a way different from (\ref{2DW}) dictated by physics. In our notation, this means using $\ddot{u}_{0/L} + 2\alpha\dot{u}_{0/L} + \omega_0^2 u_{0/L} = (\rho_m d)^{-1}\left.(p_{\rm ex} - p)\right\vert_{x=u(t,\Gamma_{0/L})}$ instead of (\ref{2DW}). From the perspective of physics, \cite{Beale2, Beale1} consider a source term to their version of (\ref{2DW}) that is different from the biophysical model \cite{ICE1,ICE2,ICE3} on which our analysis is based.


A recent study \cite{Beale3} has extended the setup considered by Beale and Rosencrans \cite{Beale2, Beale1}. In particular, interaction effects between the constituent points on the boundary are included using the Laplace-Beltrami operator considered in our equation (\ref{2DW}) as well. From the applied perspective, this setup is more realistic than Beale's harmonic oscillator model as the reacting bounding surfaces of the cavity feature interactions due to elastic coupling inside the boundary material. In the aforementioned contribution, these surfaces are termed ``non-locally reacting surfaces''. For their model, \cite{Beale3} prove the existence, uniqueness, and asymptotic stability of global solutions to the mixed problem for the wave equation of Carrier type with acoustic boundary conditions for non-locally reacting boundaries. Additionally a nonlinear impenetrability condition is considered.
 

What will be exploited here regarding the concrete situation of Fig.~\ref{iTD} is that small membrane vibrations typically are of the order of $10\,\text{nm}$ and thus $\ll L$, where $L$ is of the order of cm. The smallness of the ratio of typical membrane vibration amplitude $U$ and geometric length scale of the acoustic cavity $L$, $U/L\ll 1$, is typical of vibrational acoustic models and therefore our perturbation theory is highly relevant to many practical situations. The present method can be generalized in a straightforward manner to $>2$ membranes and $>1$ acoustic domains, mutually coupled by means of an elastic structure such as a membrane.

\paragraph{Acoustic boundary conditions} The acoustic wave equation is incomplete without specification of boundary conditions. Biologically, e.g., a lizard's eardrum cavity is bounded by a bony structure, which acts as a immobile, acoustically hard wall, as well as by eardrums at either end. The tympanic membranes $u_{0/L}$ are much lighter than the bony structure and start oscillating after an incident pressure signal appears. 

The boundary conditions for the acoustic wave equation are derived from the linearized Euler's equations, cf. \cite[chapter 2]{Temkin} for the fluid we usually call air. Starting from the Navier-Stokes equations inside the air-filled cavity $\Omega(t)$, one considers the inviscid case, i.e., where $\mu=0$. The result is Euler's equation. Performing the acoustic linearization procedure for the mass continuity equation and Euler's equation, one gets the acoustic wave equation. Restricting the linearized version of Euler's equation to the boundary, one obtains Neumann-type boundary conditions \cite[chapter 2]{Temkin},
\begin{align}
\label{boundcon}
\partial_t(\hat{\mathbf{n}} \cdot \mathbf{v})=-\dfrac{(\hat{\mathbf{n}} \cdot \nabla)p}{\rho_0}\text{ on }\partial\Omega(t).
\end{align}
Here $\hat{\mathbf{n}}$ denotes the outward unit normal to the boundary $\partial\Omega(t)$ of the cavity $\Omega(t)$, $\mathbf{v}$ is the acoustic fluid velocity, and the dot $\cdot$ in $\hat{\mathbf{n}}  \cdot \mathbf{v}$ denotes a scalar product of the vectors $\hat{\mathbf{n}}$ and $\mathbf{v}  \in \mathbb{R}^3$.

Equation (\ref{boundcon}) is a consequence of the so-called ``no-slip" boundary condition \cite[Ch.~2]{Temkin}, which requires that the speed of air at the boundary equals that of the boundary itself. Replacing $\hat{\mathbf{n}}\cdot\mathbf{v}$ by $\dot{u}_{0/L}$ in (\ref{boundcon}) is also referred to as the ``\emph{impenetrability conditions}''; cf. \cite{Beale3} using the acoustic velocity potential instead of the acoustic pressure and fully neglects both the physical origin of the boundary's variation in time and the ensuing mathematical nature embodied by Eq.~(\ref{boundcon}). Instead, Beale \cite{Beale3} used fictitious oscillators. We now spell out the details for the situation of Figs.~\ref{iTD}\,(d)\&(e).

The boundary condition in (\ref{boundcon}) decomposes into the static cylinder wall and the time-varying end caps of $\Omega(t)$, the eardrums that respond to sound. In cylindrical coordinates \cite{ICE2}, $\partial_r p(t,r,\phi,x)=0$ as the wall does not react to the acoustic pressure. Physically, this means that the cavity wall is idealized as infinitely heavy, a reasonable assumption. Hence, the cavity wall's speed vanishes, i.e., $\mathbf{v} \equiv 0$, and we obtain the traditional Neumann boundary condition with vanishing normal derivative of $p$. On the end caps, however, $\partial_{x} p(t,y,z,x=u_0(t,y,z) = -\rho_0\partial_t u_0(t,y,z)$ on $x=u_0(t,y,z)$ and $\partial_{x} p(t,y,z,x=L+u_L(t,y,z) = -\rho_0\partial_t u_L(t,y,z)$ on $x=L + u_0(t,y,z)$. On the parts of the end caps that do not undulate, that is, for $(y,z,x)\in D^2_{a_{\rm cyl}}\times\lbrace 0,L\rbrace\setminus (\Gamma_0\cup\Gamma_L)$, (\ref{boundcon}) reduces to $\partial_x p(t,r,\phi,x=0)=0$ and $\partial_x p(t,r,\phi,x=L)=0$ in cylindrical coordinates. Explicitly, the parts of the end caps in question are defined by the constraints $a_{\rm tymp} < r < a_{\rm cyl}$ and $\phi\in (0,\beta)\cup(2\pi - \beta, 2\pi)$ in cylindrical polar coordinates.

The membranes $u_{0/L}$ satisfy Dirichlet boundary conditions \cite{ICE2} on the boundaries of what is taken to be a circular sector $\Gamma_{0/L}$ for left and right eardrums at $x=0$ and $x=L$, as shown in Fig.~\ref{iTD}\,(e): $u_{0/L}=0$ on $\partial\Gamma_{0/L}$, a circle of  radius  $a_{\rm tymp}$.

\section{Geometric perturbations}
\label{gppo}

The subtlety in the treatment of a model such as ICE is that the geometry $\Omega(t)$ of the acoustic wave equation for the pressure $p$ implicitly depends on the membranes as part of the boundary $\partial\Omega(t)$, while the dynamics of the membranes as 2-dimensional manifolds contains $p$ as source term; cf. Eqs. (\ref{3DW}), (\ref{2DW}), and (\ref{boundcon}). That is, at $t>0$ the boundary $\partial\Omega(t)\supseteq u_{0/L}(t,\Gamma_{0/L})\equiv\lbrace(\mathbf{y},0/L+u_{0/L}(t,\mathbf{y}))\in\mathbb{R}^3\vert \mathbf{y}\in \Gamma_{0/L}\rbrace$
contains the membranes' hypersurfaces implicitly in the boundary conditions (\ref{boundcon}) and explicitly in the Laplacian $\Delta_t$. In turn, the membranes' hypersurfaces depend on the pressure difference $[p_{\rm{ex}}(t)-p(t)]$ as indicated in (\ref{2DW}). How to approach these two subtleties and solve the coupled dynamics of $p$ and $u_{0/L}$? Experimentally, typical membrane vibration amplitudes $U\lesssim 10\,\text{nm}$ are much smaller than the characteristic axial length scale $L\simeq 2\,\text{cm}$ of the cavity, i.e., $U/L \simeq 10^{-6}\lll 1$, so that a perturbation theory ought to be designed to reduce the acoustic wave equation on $\Omega(t)$ to a reference acoustic wave equation on $\Omega_0$. Analogously to Dirac's perturbation theory originating from quantum theory \cite{Dirac2}, the Laplacian $\Delta_t$ is decomposed into a Laplacian $\Delta$ on the unperturbed domain $\Omega_0$ and a time-dependent perturbation operator $\mathsf{V}$ to account for the time-dependent deviation of $\Omega(t)$ from the stationary reference cylinder $\Omega_0$. We therefore turn to the underlying ideas. 

\subsection{Time-dependent perturbation theory in the style of Dirac}
\label{dirac_style}

In the late twenties of last century, Paul Dirac devised a time-dependent perturbation theory in the context of quantum mechanics \cite{Dirac1,Dirac2}. See Reed and Simon \cite{Reed1, Reed2, Reed4} for a lucid description and first mathematical proofs. Here we describe its main idea from the perspective of Dirac's classic on quantum mechanics \cite[\S\S44-46]{Dirac2} but along the lines of Duhamel's principle \cite[\S XIV.5]{Yosida}, which turns out to be most suitable to the present approach. 

We start with the time-dependent Schr\"{o}dinger equation
\begin{align}
\label{timedeppsde}
i\hbar\partial_t\psi = \left[\mathsf{H}_0+\mathsf{V}(t)\right]\psi ,
\end{align}
where $\partial_t = \partial/\partial t$ denotes the partial derivative with respect to time $t$, $\hbar$ is Planck's constant, and $\mathsf{H}_0=-(\hbar^2/2m)\Delta$ is proportional to the operator $-\Delta$ with proportionality constant $\hbar^2/2m$ while $\mathsf{V}(t)$ is a time-dependent perturbation. We assume the operators $\mathsf{H}_0$ and $\mathsf{V}(t)$ to be defined on a single, simply-connected, compact domain $\Omega_0$ in $\mathbb{R}^3$ for all $t\geq 0$ The linear operator $\mathsf{H}=\mathsf{H}_0+\mathsf{V}(t)$ is what is called a Hamiltonian system in quantum mechanics. Here the self-adjoint $\mathsf{H}_0$ is known completely in the sense that its eigenfunctions $\lbrace\Psi^{(0)}_n\rbrace_{n\in\mathbb{N}}$ and eigenvalues $\lbrace\lambda_n^{(0)}\rbrace_{n\in\mathbb{N}}$ are supposed to be known so that a spectral representation \cite[\S VI.5]{Kato} is well-defined. In the present case, we treat a spatial domain that is so bounded that the statement is evident.

Let us call $\mathsf{V}\psi\equiv f$ and use it as a source term in
\begin{align}\label{sourcesde}
(i\hbar\partial_t-\mathsf{H}_0)\psi = f.
\end{align}
With the semi-group $\mathsf{U}_0(t,\tau)=\exp\left[-i\hbar^{-1}(t-\tau)\mathsf{H}_0\right]$ from, say, Stone's theorem, the solution can be rewritten in Duhamel representation \cite[\S XIV.5]{Yosida}, as one verifies by direct differentiation,
\begin{align}
\label{duhamelzeug}
\psi(t)=\mathsf{U}_0(t,0)\psi(0)+\int_0^t d\tau\,\mathsf{U}_0(t,\tau)f(\tau)
\end{align}
so that we get an integral representation for our original problem,
\begin{align}\label{diraczeug}
\psi(t)=\mathsf{U}_0(t,0)\psi(0)+\int_0^t d\tau\,\mathsf{U}_0(t,\tau)\mathsf{V}(\tau)\psi(\tau).
\end{align}
In quantum mechanics, one takes a judicious choice for $\psi(\tau)$ in the right-hand  side of (\ref{diraczeug}), such as $\psi(\tau)=\psi(0)$, and iterates (\ref{diraczeug}). Here we shall do something similar by taking due advantage of the underlying physics of our time-varying boundary conditions, which apparently has been overlooked until now.

Three remarks are in order. First, in physics literature it is common usage to put $\tilde{\psi}(t)\equiv\mathsf{U}_0(-t,0)\psi(t)$ so as to arrive at the ``interaction picture'' also introduced by Dirac \cite[\S44]{Dirac2},
\begin{align}\label{interdirac}
\tilde{\psi}(t)=\psi(0)+\int_0^t d\tau\,\tilde{\mathsf{V}}(\tau)\tilde{\psi}(\tau)
\end{align}
where $\tilde{\mathsf{V}}(t)=\mathsf{U}_0(-t,0)\mathsf{V}(t)\mathsf{U}_0(t,0)$ is unitarily equivalent to $\mathsf{V}(t)$. Once $\mathsf{V}\equiv\mathsf{0}$, $\tilde{\psi}(t)=\psi(0)$. Second, except for Dirac nearly all quantum mechanics books work coordinate-wise through the expansion $\tilde{\psi}(t)=\sum_n c_n(t)\Psi^{(0)}_n$ and solve for $c_n(t)$. Here, however, we stick to the Duhamel representation (\ref{duhamelzeug}). Third, and important, starting with $\psi^{(0)}(t)=\psi(0)$, one can iterate (\ref{interdirac}) and obtain higher-order solutions.
In the present context we substitute, so to speak, $t\to i\hbar t$ and $\mathbf{x}\to (\hbar /\sqrt{2m})\mathbf{x}$ as well as $\mathsf{V}\to -\mathsf{V}$ in (\ref{timedeppsde}), and study
\begin{align}
\label{struc}
\partial_t\psi = \left[\Delta + \mathsf{V}(t)\right]\psi
\end{align}
We now let $\psi$ become a vector-valued function $\bm{\psi}=(\psi_{1},\psi_2,...,\psi_n)$ with components $\psi_i,\,1\leq i\leq n$. Instead of the scalar operators $\Delta, \mathsf{V}$, we shift to quadratic matrices with operators as entries. That is, we analyze an evolution equation of the form
\begin{align}
\label{struc2}
\partial_t\bm{\psi}=\left[\mathbf{\Delta}+\mathbf{\mathsf{V}}(t)\right]\bm{\psi}.
\end{align}
As a first step, we recast the model equations (\ref{3DW}) \& (\ref{2DW}) into an equation that has the evolution equation structure as indicated by (\ref{struc2}). Our original problem stems from the wave equation $(\partial_t^2-\Delta_t)p=0$ as in (\ref{timedeppsde}) \& (\ref{sourcesde}) but with $c^2=1$. Through the simple substitution $\partial_t p = q$, we can reduce the second-order wave equation to a first-order differential equation with the structure of (\ref{struc2}). The boundary condition to the acoustic wave equation generates the coupling with the damped wave equation for both eardrums, which constitute a time-varying boundary condition for the original wave equation. That is, we are going to study a problem of the following shape,
\begin{align}
\label{struc3}
\partial_t^2 p =\Delta_t p \Rightarrow \partial_t\bm{\psi} = \left[\mathbf{\Delta}+\kappa (\mathds{1}_{\delta\Omega}\mathbf{\Delta}\mathds{1}_{\delta\Omega}) \right]\bm{\psi} = \left[\mathbf{\Delta}+\mathsf{V}(t)\right]\bm{\psi}
\end{align}
where $\bm{\psi}=(p,\partial_tp)$ and $\mathsf{V}(t)$ is a matrix of operators taking the role of a time-dependent perturbation in the style of Dirac \cite[\S\S44-46]{Dirac2}. In the above equation (\ref{struc3}), $\kappa=\kappa(t,\mathbf{x})=\pm 1$ describes the local increase or decrease of the time-varying domain $\Omega(t)$, an open set, with respect to the unperturbed domain $\Omega_0$, and $\delta\Omega(t)$ indicates the symmetric difference $\delta\Omega(t)\equiv\Omega_0\ominus\Omega(t)$ of the sets $\Omega_0$ and $\Omega(t)$ with indicator function $\mathds{1}_{\delta\Omega(t)}$. The operator 
\begin{equation}
  (\mathds{1}_{\delta\Omega}\mathbf{\Delta}\mathds{1}_{\delta\Omega}) \equiv \Delta\, |_{\delta\Omega(t)}
\label{struc4}
\end{equation}
is the Laplacian restricted to the fluctuating domain $\delta\Omega(t)$, a (separate unit) that -- despite a slight misuse of language -- effectively does what it formally looks like. In this way, we can take into account the physical variation of the cavity domain $\Omega(t)$ as a consequence of the eardrum or whatever membrane variations, which are always very small, and the right-hand side of (\ref{struc3}) then leads directly to Duhamel \cite[\S XIV.5]{Yosida}. 

As compared to its quantum-mechanical origin, where time evolution is unitary and the unperturbed action is generated by a unitary semigroup, the present perturbation theory is based on an unperturbed action generated by a contraction instead of unitary semigroup, which simplifies the ensuing arguments substantially. 

\subsection{Intermezzo: The need for perturbation theory}
\label{intermezzino}
Since dynamical problems associated with ICE have been scarcely studied,  it is worthwhile to quickly analyze the appropriate setup in the simplest possible context. 
To this end, we consider a modification of a problem discussed by Shearer and Levy \cite[\S6.2.2]{sh+l}, viz., the inhomogeneous time-dependent boundary-value problem
\begin{equation}
\partial_t^2 w - \partial_x^2 w = f(t,x) \quad \mathrm{with} \quad t>0,x\in[0,\pi]
\label{u-example0}
\end{equation}
where $w = w(t,x)$. The initial conditions are $w(0,x)=0$ and $\partial_t w(0,x)=0$ for $0<  x <\pi$ and the time-dependent boundary conditions are given by $\partial_x w(t,0)=b_0(t)$ and $\partial_xw(t,\pi)=b_\pi(t)$ for $t>0$, with $b_0$ and $b_\pi$ sufficiently smooth but nonzero. We note the boundaries at $x=0$ and $x=\pi$ are still fixed. 

For Neumann boundary conditions  $\partial_x w(t,0) = \partial_xw(t,\pi)=0$, we know $-\partial_x^2$ is a non-negative self-adjoint operator on $[0,\pi]$ with $(\sqrt{2/\pi})\cos(nx)$ and $n$ positive integer as a complete orthonormal set of eigenfunctions. We now make the separation ansatz $w(t,x) = \sum_{n \ge 0} w_n(t)\, (\sqrt{2/\pi})\cos(nx)$. Obviously, $\partial_x\cos(nx)=0$ if evaluated at $x = 0$ and $x=\pi$. So the correct way to understand the boundary conditions is: $\lim_{x\to 0^+}\partial_x w(t,x) = b_0(x)$ and $\lim_{x\to \pi^-}\partial_x w(t,x) = b_\pi(x)$. That is, the  boundary is always approached \emph{from the inside} of the domain and convergence of the infinite series in the above ansatz is not uniform but point-wise in the open interval, here associated with the Gibbs phenomenon \cite[\S7.5.2]{sh+l}. 

A decently straight calculation gives, with $f_n$ as $n$-th Fourier component of $f$,
\begin{equation}
\ddot{w}_n(t) = -n^2 w_n(t) + \frac{2}{\pi}\left[b_0(t)+\cos(n \pi)b_\pi(t)\right]+f_n(t) \equiv -n^2 w_n(t)+F_n(t) .
\label{u-example1}
\end{equation}
Using the above convention of always approaching the boundary from the inside, one easily verifies that $F_n$ can also be written as the n-th component of $f$ plus two surface $\delta$-functions and, without components $n$ but back at PDE level, 
\begin{equation}
\partial_t^2 w(t,x) - \partial_x^2 w = F(t,x) = f(t,x)+[\delta(x) b_0(t)+\delta(x-\pi) b_\pi(t)]
\label{u-example2}
\end{equation}
with $\partial_x w(t,0)=0=\partial_xw(t,\pi)$. By definition, the surface-$\delta$ only requires semi-continuity from the inside. Indeed, this is an acceptable assumption given that even the technique of partial Fourier expansion built into Green's function and PDE treatments via eigenfunction expansions requires it. 

Duhamel reproduces the standard solution to (\ref{u-example1}) and directly solves (\ref{u-example2}) on the operator level, without specifying the component $n$. Even the more so, in ICE the end points are boundary points \emph{and} functions of time. To get explicit solutions, we need a systematic perturbation theory, which we now turn to. 

\subsection{Perturbative approach to acoustic geometries}
\label{patag}

The wave equation
\begin{align}
\dfrac{\partial^2 p}{\partial t^2}-c^2\Delta_t p &=0
\end{align}
with the boundary conditions (\ref{boundcon}), i.e., 
\begin{align}
(\hat{\mathbf{n}} \cdot \nabla)p=-\rho_0\partial_t(\hat{\mathbf{n}} \cdot \mathbf{v})\text{ on }\partial\Omega(t),
\end{align}
shall be solved by a Dirac-inspired perturbation theory, as in (\ref{struc}). In order to reduce the partial differential equation from the domain $\Omega(t)$ to the domain $\Omega_0,$ we define the perturbation operator $\mathsf{V}$ by
\begin{align}
\Delta_t &= \mathds{1}_{\Omega(t)}\Delta\mathds{1}_{\Omega(t)} =\mathds{1}_{\Omega_0}\Delta\mathds{1}_{\Omega_0}+\kappa \, \mathds{1}_{\delta\Omega}\Delta\mathds{1}_{\delta\Omega}\nonumber \\
&=\Delta +\kappa \, (\mathds{1}_{\delta\Omega}\Delta\mathds{1}_{\delta\Omega}) \equiv \Delta+\mathsf{V}(t) 
\label{defV}
\end{align}
where $\mathsf{V}(t)$ is specified through (\ref{struc3}),  (\ref{struc4}), and (\ref{defV}), and for the sake of completeness $\mathds{1}_X$ denotes the indicator function of a Borel set $X\subset \mathbb{R}^3$, 
\begin{align}
\label{pertop2}
\mathds{1}_X = \left\lbrace\begin{array}{cc}1 & \mathrm{for} \quad \mathbf{x}\in X\\ 0 & \!\! \mathrm{for} \quad  \mathbf{x}\in X^\complement \end{array}\right. 
\end{align}
where $X^\complement$ is the complement of $X$ as subset of whatever. Furthermore, $\kappa$ takes the value $+1$ if $\mathbf{x}\in\Omega(t)$ and $\mathbf{x}\not\in\Omega_0$, and the value $-1$ if $\mathbf{x}\in\Omega_0$ and $\mathbf{x}\not\in\Omega(t)$. This accounts for the observation that the deformation of $\Omega_0$ by membrane vibrations can either increase or decrease the volume of $\Omega(t)$, and the $\pm$ of $\kappa$ tells us so. The fluctuation term $\mathsf{V}(t) = \kappa \, (\mathds{1}_{\delta\Omega}\Delta\mathds{1}_{\delta\Omega})$ has a spatial extent that is small (viz., nm) as compared to the size of the cavity (viz., cm). Nevertheless it facilitates hearing and sound localization. 

The Laplacian $\Delta_t$ in (\ref{defV}) is to be defined as operator on a suitable domain of continuous and sufficiently oft differentiable functions \cite{Yosida} in the Hilbert space $L^2(\Omega_{\rm max})$. Here $\Omega_{\rm max}$ is a stationary, regular, compact region in $\mathbb{R}^3$ such that the subregion $\Omega_t$ with $t \ge 0$ and undulating boundary $\partial\Omega_t$ satisfies $\Omega_t\subset \Omega_{\rm max}$ for all $t\ge 0$. The existence of $\Omega_{\rm max}$ is guaranteed by the dissipative nature of the underlying dynamics; more in particular, by $\alpha > 0$ in (\ref{2DW}).
Although $\text{Dom}(\Delta_t)$ requires functions to be defined only for arguments $\mathbf{x} \in \Omega_t$, we demand them to be restrictions of functions defined on $\Omega_{\rm max}$ that live in a suitable subspace $V\subseteq H^1(\Omega_{\rm max})$.
The boundary $\partial\Omega_t$ being a two-dimensional manifold in $\mathbb{R}^3$ and, hence, having Lebesgue measure zero in $\Omega_{\rm max}$, the discontinuities just mentioned do not count in $L^2(\Omega_{\rm max})$. The ``unperturbed'' operator $\Delta = \Delta_{t=0}$ with Neumann boundary conditions on $\partial\Omega_0$ is (essentially) self-adjoint \cite{Yosida}. 

As explained above in section~\ref{dirac_style}, we choose the ansatz $p(t,\mathbf{x})= p^{(0)}(t,\mathbf{x})+p^{(1)}(t,\mathbf{x})$ with $p^{(0)}$ as the solution to the zeroth-order equation and $p^{(1)}$ as the solution to the first-order equation. As  boundary condition of the zeroth-order problem we put 
\begin{align}
\label{neumann_11}
(\hat{\mathbf{n}} \cdot \nabla)p^{(0)} = 0\text{ on }\partial\Omega_0,
\end{align}
which is Neumann. For the first-order problem, we then get
\begin{align}
\label{neumann_12}
(\hat{\mathbf{n}} \cdot \nabla)p^{(1)}=-\rho_0\partial_t(\hat{\mathbf{n}} \cdot \mathbf{v})\text{ on }\partial\Omega(t).
\end{align}
The initial conditions to the zeroth and first-order problem are obtained on physical grounds. For instance, lizards perceive sound as the superposition of the external stimulus $p_{\rm ext}$ and the internal pressure $p$ operating on their eardrums. Lizards and other terrestrial vertebrates need to azimuthally localize a sound source in their environment through neuronally determining the resulting \emph{internal} time difference (iTD). The difference between the incident pressure signal $p_{\rm{ex}}$, which in turn generates a response pressure $p$ in the eardrum cavity as the resultant of the interaction between both eardrums, causes the vibrations of the membranes $u_{0/L}$ of either ear and gives rise to the iTD, which in general differs from the external time difference ITD. 

The no-slip boundary condition \cite[chapter 2]{Temkin} of the zeroth-order problem is  independent of any membrane vibration and such that $p^{(0)}(t=0)=0=\partial_tp^{(0)}(t=0)$. Analogously to the quantum-mechanical Dirac perturbation theory, we regard the first-order contribution $p^{(1)}$ to the cavity pressure as a response to the perturbation such that $p^{(1)}(t=0)=0=\partial_tp^{(1)}(t=0)$. Thus, in view of (\ref{defV}), the perturbation ansatz for the acoustic wave equation in (\ref{3DW}) results in
\begin{align}
\mathcal{O}(1)&: \dfrac{\partial^2 p^{(0)}}{\partial t^2}-c^2\Delta p^{(0)} =0, \\
\mathcal{O}(U/L)&: \dfrac{\partial^2 p^{(1)}}{\partial t^2}-c^2\Delta p^{(1)} = c^2\mathsf{V} p^{(0)}.
\end{align}
Since both initial and boundary conditions for the zeroth-order equation vanish, $p^{(0)}(t,\mathbf{x}) = 0$, i.e., $p(t,\mathbf{x})=p^{(1)}(t,\mathbf{x})$. As a difference of Laplacians, $\mathsf{V}=\kappa (\mathds{1}_{\delta\Omega}\Delta\mathds{1}_{\delta\Omega})$ is a linear operator (on $p$) such that $0\in\text{Ker}[\mathsf{V}]$. Since $p^{(0)}(t,\mathbf{x})=0$, it follows $\mathsf{V}[p^{(0)}]=0$. Thus, the first-order problem simplifies to 
\begin{align}
\label{acfirst}
\mathcal{O}(U/L)&: \dfrac{\partial^2 p^{(1)}}{\partial t^2}-c^2\Delta p^{(1)} = 0.
\end{align}
We note that instead of $\Delta_t$ the Laplacian $\Delta$ on the stationary domain occurs in the above equation (\ref{acfirst}).
It is good to realize that appearances are deceiving in that $p^{(1)} \equiv 0$ is a solution to (\ref{acfirst}) but does not satisfy the boundary condition (\ref{neumann_12}), a problem to which we return in (\ref{dummyeq2}).

As detailed in Section~\ref{dirac_style}, we solve the eigenvalue problem for the 
zeroth-order operator $\Delta$ on $\Omega_0$ subject to the homogeneous, Neumann, boundary conditions and obtain the eigenfunctions $\Psi_{\mathbf{n}}$ and eigenvalues $\lambda_{\mathbf{n}} \ge 0$. Since $\Omega_0$ is a compact simply-connected domain, the Laplacian has a discrete set of eigenfunctions and eigenvalues \cite{Yosida}. They follow from
\begin{align}\label{eigenzit}
\Delta\Psi_{\mathbf{n}} &= -\lambda_\mathbf{n}\Psi_{\mathbf{n}}\text{ on }\Omega_0\quad ,\quad (\hat{\mathbf{n}} \cdot \nabla)\Psi_{\mathbf{n}} = 0\text{ on }\partial\Omega_0.
\end{align}
Up to a normalization constant $\mathcal{N}_\mathfrak{n}^{-1}$, the solution to the eigenvalue equation is, after separation of variables in cylindrical coordinates, given by
\begin{align}
\Psi_{\mathbf{n}}(\mathbf{x})&=\mathcal{N}^{-1}_{\mathbf{n}}J_{n_2}\left(\dfrac{\mu_{n_1,n_2}r}{a_{\rm cyl}}\right)e^{in_2\phi}\cos\left(\dfrac{n\pi x}{L}\right),\\
\lambda_{\mathbf{n}}&=\dfrac{\mu_{n_1,n_2}^2}{a_{\rm{cyl}}^2}+\dfrac{n^2\pi^2}{L^2}
\end{align}
where $\mu_{n_1,n_2}$ denotes the $n_1$-st non-negative extremum of the Bessel function $J_{n_2}$ with $\mu_{n_1,n_2}\in\lbrace x\vert J'_{n_2}(x)=0\rbrace$. This concludes the preparation of tackling the perturbed problem. 

In order to solve the wave equation with inhomogeneous boundary conditions, we use an as yet not quite standard technique of translating 
inhomogeneous boundary conditions into homogeneous boundary conditions with a corresponding source term; see \cite{Pan1, Pan2} for an experimental study of this topic. As in Vossen et al.~\cite{ICE1}, we extend the eigenfunctions $\Psi_{\mathbf{n}}$ due to their smoothness -- and to Neumann with error $\mathcal{O}((U/L)^2)$ -- from $\Omega_0$ to $\Omega_0\cup\Omega(t)$ so that
\begin{align}
\label{dummyeq}
\mathcal{O}(U/L): \dfrac{\partial^2 p^{(1)}}{\partial t^2} - c^2\Delta p^{(1)} &= -\rho_0c^2\partial_t\mathbf{v_n}\delta(\mathbf{y}\in\partial\Omega(t)) \ .
\end{align}
$\mathbf{v}_{\mathbf{n}}$ denotes the normal component of the acoustic velocity $\mathbf{v}$ of air at the boundary $\partial\Omega(t)$ and $\delta(\mathbf{y}\in\partial\Omega(t))$ is, as before, a surface-Dirac-delta distribution. We define $\delta(\mathbf{y}\in\Gamma(t))\equiv -\mathbf{n}_{\mathbf{y}}\nabla\mathds{1}_{\Omega(t)\setminus(\partial\Omega(t)\setminus\Gamma(t))}(\mathbf{y})$ for the subdomain $\Gamma(t)\subset\Omega(t)$; cf. Lange \cite{surfdelta} for a definition in terms of the indicator function for the entire boundary. 

The usage of Dirac distributions as source functions, as preferred in the mathematical physics literature, is an equivalent way of writing the usual boundary terms in a weak formulation \cite{Evans,sh+l} of the problem posed in a space of distributions. On the other hand, the above procedure is more direct and even supported by textbook presentations \cite[\S~6.2.2]{sh+l} that, as one can easily verify (see also below), present an algorithm equivalent to (\ref{dummyeq}). 

Practically, the surface-Dirac-delta distribution (not meant to be a distribution in the sense of Schwartz \cite{schwartz}) is a measure concentrated on a two-dimensional manifold, the tympanic membrane, with the following property,
\begin{align}
\int_{\Omega(t)}d^3\mathbf{x}\,f(\mathbf{x})\delta(\mathbf{x}\in\Gamma(t))=\int_{\Gamma(t)\subseteq\partial\Omega(t)}d^2\mathbf{y}\,f(\mathbf{y}) 
\end{align}
where we denote points on the boundaries $\partial\Omega(t)$ by $\mathbf{y} \in \partial\Omega(t)$  and  points in the $t$-dependent volume $\Omega(t)$ by $\mathbf{x}\in\Omega(t)$. The source term in (\ref{dummyeq}) can be expressed in terms of the membrane accelerations $\partial_t^2 u_{0}$ for the left and $\partial_t^2 u_L$ for the right membrane. The no-slip condition \cite[chapter 2]{Temkin} tells us that we have $\mathbf{v}_n(t,u_{0/L}(t,\mathbf{y})) \simeq \partial_t u_{0/L}(t,\mathbf{y})$ for weakly curved tympanic membranes (here nm amplitudes for a cm-large eardrum), where we can neglect curvature effects of the membranes. 

In the ICE model, the membrane amplitudes $u_0 \, \& \, u_L$, or for short $u_{0/L}$, are the only mobile elements of the boundary $\partial\Omega(t)$ such that (\ref{dummyeq}) reads in terms of the membrane accelerations $\partial_t^2 u_0$ and $\partial_t^2 u_L$ in the source term
\begin{align}
\label{dummyeq2}
\dfrac{\partial^2 p^{(1)}}{\partial t^2} - c^2\Delta p^{(1)} &= \rho_0c^2\,[(\partial^2_tu_0)\delta(\mathbf{x}\in u_0(t,\Gamma_0))+(\partial^2_tu_L)\delta(\mathbf{x}\in u_L(t,\Gamma_L)] 
\end{align}
where we have ignored additional contributions to the normal velocity $\mathbf{v_n}=\hat{\mathbf{n}\mathbf{v}}$ of order $\sim U/L$, which stem from the tympanic membranes' curvature, in order to stay in first-order perturbation theory. Abbreviating the right-hand side of (\ref{dummyeq2}) by $q=q(t,\mathbf{x})$, the equation takes the form
\begin{align}
\dfrac{\partial^2 p^{(1)}}{\partial t^2} - c^2\Delta p^{(1)} &= q,
\end{align}
As in subsection \ref{dirac_style}, the Duhamel principle \cite[\S XIV.5]{Yosida} (see also \cite[\S\S 5.9-5.11]{Zeidler2}) can now be applied by noting that the above equation is equivalent to the first-order differential equation system,
\begin{align}\label{dummyeq3}
\dfrac{\partial}{\partial t}\left(\begin{array}{c}p^{(1)} \\ \partial_t p^{(1)}\end{array}\right) &= \left(\begin{array}{cc}0 & 1\\ c^2\Delta & 0\end{array}\right)\left(\begin{array}{c}p^{(1)} \\ \partial_t p^{(1)}\end{array}\right) + \left(\begin{array}{c}0 \\ q\end{array}\right).
\end{align}
On the basis of semigroup theory -- cf. Reed and Simon \cite[chapter 10]{Reed2} and Yosida \cite[Chapter XIV]{Yosida} -- we use exponentiation and the initial conditions $p^{(1)}(t=0)=0=\partial_t p^{(1)}(t=0)$ to solve the linear system (\ref{dummyeq3}) of first-order differential equations and substitute back the definition of the source term $q$ so as to obtain an explicit expression in terms of operator-sine functions; cf. \cite{Opsincos,Opsincos2} or Zeidler's textbook presentation \cite[\S\S 5.9-5.11]{Zeidler2}. We define the operator Green's function
\begin{align}
\mathsf{G}_{\Delta}(t-\tau)\equiv\sqrt{-c^2\Delta}^{-1}\sin\left((t-\tau)\sqrt{-c^2\Delta}\right)
\end{align}
and the corresponding spectral Green's functions by means of projection
\begin{align}
G_{\mathbf{n}}(t-\tau)\equiv\langle\Psi_{\mathbf{n}}\vert G_{\Delta}(t-\tau)\vert\Psi_{\mathbf{n}}\rangle_{\Omega(t=0)}
\end{align}
where the $\lbrace\Psi_{\mathbf{n}}\vert\mathbf{n}\in\mathbb{N}^3_0\rbrace$ denote the eigenfunctions from (\ref{eigenzit}). We notice that $G_{\Delta}=\sum_{\mathbf{n}}\vert\Psi_{\mathbf{n}}\rangle G_{\mathbf{n}}(t-\tau)\langle\Psi_{\mathbf{n}}\vert$ by a resolution of the identity thanks to completeness of the eigenfunction set $\lbrace\Psi_{\mathbf{n}}\vert\mathbf{n}\in\mathbb{N}^3_0\rbrace$. The solution of the first-order differential equation system (\ref{dummyeq}) is then given by
\begin{align}
&p^{(1)}(t,\mathbf{x})=\int_{0}^{t}d\tau\,\mathsf{G}_{\Delta}(t-\tau)q(\tau,\mathbf{x})\\
&=\rho_0 c^2\sum_{i\in\lbrace 0,L\rbrace}\sum_{\mathbf{n}}\Psi_{\mathbf{n}}(\mathbf{x})\int_{0}^{t}d\tau\,G_{\mathbf{n}}(t-\tau) \int_{u_i(\tau,\Gamma_i)}dS_i(\tau)\, \bar{\Psi}_{\mathbf{n}}(\mathbf{x})\partial_t^2 u_i\delta(\mathbf{x}\in u_i(\tau,\Gamma_i)).
\end{align}
The surface elements $dS_i, i\in\lbrace 0,L\rbrace$ are time-dependent for time-varying hypersurfacse $\text{graph}_{\Gamma_i} u_i\equiv\lbrace (x,\mathbf{y}):x=u_i(t,\mathbf{y}), \mathbf{y}\in\Gamma_i\rbrace$. Explicitly, a surface element $dS_i$ is given for $i\in\lbrace 0,L\rbrace$ by
\begin{align}
dS_i(t) = dS_i\sqrt{1+(\nabla u_i)^2}\simeq dS_i,
\end{align}
where $dS_i$ denotes the surface element of $\Gamma_i$ and we can neglect higher-order membrane curvature contributions due to the smallness of the vibration amplitudes: $U\simeq 10\,\text{nm}$ vs $L\simeq a_{\rm{cyl}}\simeq a_{\rm{tymp}}\simeq 1\,\text{cm}$ so that $U/L\simeq 10^{-6}$. The surface integral then simplifies to an integral over the boundary $\partial\Omega_0\supset\Gamma_0,\Gamma_L$ of the stationary domain $\Omega_0$,
\begin{align}
p^{(1)}(t,\mathbf{x})&=-\rho_0 c^2\sum_{i\in\lbrace 0,L\rbrace}\sum_{\mathbf{n}}\Psi_{\mathbf{n}}(\mathbf{x})\int_{0}^{t}d\tau\,G_{\mathbf{n}}(t-\tau) \int_{\Gamma_i}dS_i\, \bar{\Psi}_{\mathbf{n}}(\mathbf{x})\partial_t^2 u_i\mathds{1}_{\Gamma_i}.
\end{align}

Setting $p= p^{(1)}$ in first-order perturbation theory, we can formulate the problem for the cavity pressure $p$ as 
\begin{align}
\label{sumeq1}
\dfrac{\partial^2 p}{\partial t^2}-c^2\Delta p = \rho_0 c^2 \, [\partial_t^2 u_0\delta(\mathbf{y}\in\Gamma_0) + \partial_t^2 u_L\delta(\mathbf{y}\in\Gamma_L)],
\end{align}
which is geometrically situated solely in the stationary domain $\Omega_0$ instead of the time-varying $\Omega(t)$. We note that the dynamics is now ``stored'' solely in the membrane displacements $u_0$ and $u_L$ and no longer implicitly in the geometry.

To make contact with the coupled ICE equations, we apply the perturbation ansatz $p=p^{(0)}+p^{(1)}$ to the membrane equations [for $u_{0/L}$, see Eq.~\ref{dummyeq2}],
\begin{align}
\label{sumeq2}
\dfrac{\partial^2 u_{0/L}}{\partial t^2}+2\alpha\dfrac{\partial u_{0/L}}{\partial t}-c^2_m\Delta_2 u_{0/L}&=\left.\dfrac{p_{\rm{ex}}-p}{\rho_m d}\right\vert_{u_{0/L}(t,\Gamma_{0/L})},
\end{align}
which are inhomogeneous second-order hyperbolic differential equations in two spatial variables. We now observe that, in first order in acoustically small quantities, the source term in the differential equations for $u_{0/L}$ is to be evaluated at $x=0/L$, where we neglect contributions of quadratic order in acoustic quantities. 

Furthermore, in lowest non-trivial order of perturbation theory, $p=p^{(0)}=0$, i.e., only the external pressure signal $p_{\rm ex}$ drives the membranes. This is in agreement with the biophysical reasoning that, say, a lizard shall locate predator or prey by membrane vibration due to an external sound source emitting the pressure signal $p_{\rm{ex}}$. Using the above two observations, the lowest-order contribution to the membrane equations consistent with the previously introduced perturbation theory is given by
\begin{align}
\dfrac{\partial^2 u^{(0)}_{0/L}}{\partial t^2}+2\alpha\dfrac{\partial u^{(0)}_{0/L}}{\partial t}-c^2_m\Delta_2 u^{(0)}_{0/L}&=\left.\dfrac{p_{\rm ex}}{\rho_m d}\right\vert_{x=0/L}\text{ on }\Gamma_{0/L}\subset\partial\Omega_0 .
\end{align}
Together with the wave equation for the cavity pressure $p$, this has the effect of decoupling the ICE equations in lowest-order perturbation theory. These equations serve as a starting point for the investigation of the validity of the piston approximation  \cite{david2, ICE2, ICE3} in subsection~\ref{apple1}. 

\subsection{Intermediate summary}

What has been achieved until now and what is its use? First and foremost, a perturbative treatment of vibrational acoustics boundary value problems has been established. Because of the vibrations of structural elements in the boundary of an acoustic enclosure, i.e., membranes in this article, the domain $\Omega(t)$ of the Laplacian $\Delta_t$ is no longer stationary but changes in time. Typically, the vibration amplitudes $U$ of the structural elements, viz., membranes, are very small compared to a characteristic length scale of the acoustic enclosure. That is, $10$\,\,nm vs. $1$\,\,cm. For a cylinder with membranes (of the eardrums) as end-caps, it makes physically sense to choose its length $L$ as a geometric length scale and to compare it with the vibration amplitudes $U$ of the membranes. 

Since the deviations of the domain $\Omega=\Omega(t)$ from the stationary $\Omega_0$, where the acoustic wave equation can be solved analytically, are very small, a Dirac-type \cite{Dirac1,Dirac2} perturbation theory argument demonstrates that we can in first-order perturbation theory reduce the boundary-value problem on the time-varying domain $\Omega(t)$ to a boundary-value problem on the stationary $\Omega_0$. In this way we exploit the underlying physics to mathematically formulate a time-dependent perturbation problem that can be solved iteratively. 

In first-order perturbation theory we have $p=p^{(0)}+p^{(1)}$, where $p^{(0)}$ has been argued to be $0$. For small-vibration amplitudes, the membranes are almost flat and corrections are of third order in derivatives of the membrane displacement, viz. $(\partial u)^3$, so that the acoustic pressure can be expressed as the solution to the acoustic wave equation on a stationary domain $\Omega_0$ with Neumann boundary conditions, identifying the normal fluid velocity with the membrane displacements, $\mathbf{n}\cdot\mathbf{v}=\partial_t u_{0}\delta(\mathbf{y}\in\Gamma_0)+\partial_t u_{L}\delta(\mathbf{y}\in\Gamma_L)$.

An aesthetic improvement can be accomplished by switching from the inhomogeneous boundary conditions to a source term using Duhamel's principle \cite[\S XIV.5]{Yosida}. Together with the pressure-difference driven membranes, the ICE model can be formulated in terms of three wave equations with homogeneous initial and boundary conditions for the acoustic pressure $p$ and the membrane displacements $u_{0/L}$. That is, equations (\ref{sumeq1}) and (\ref{sumeq2}). Applying the technique of Picard iterations and setting $p^{(0)}=0$ in agreement with the perturbative argument before, the three partial differential equations decouple. The membranes are driven by the external pressure only and modulated by the internal dynamics of, in our case, the interaural cavity; cf. Fig.~\ref{iTD}. The first non-vanishing contribution to the acoustic pressure $p^{(1)}$ then can be obtained from solving the inhomogeneous acoustic wave equation with source terms given by the $0$-th order membrane displacements, i.e., membranes driven by the external pressure only.

One of the major advantages of the present approach is that we can iterate the Duhamel expression systematically and obtain higher-order corrections straightforwardly, as will be demonstrated in the next Section~\ref{appl}.

\section{Applications of geometric perturbation theory}
\label{appl}

We now apply the formalism of acoustic boundary-condition dynamics (ABCD) developed in Section~\ref{gppo} so as to solve the ICE equations (\ref{3DW}) \& (\ref{2DW}). As a consequence of the solution, we can for the first time derive the piston approximation \cite{ICE2, ICE3}, which has been studied from a geometric viewpoint in full mathematical detail by Heider and van Hemmen \cite{david2} by means of a generalized cut-off criterion for non-axial cavity modes. The evanescent mode cut-off criterion referred to as \emph{modal cut-off criterion} in the following is generalized and reproduced by an exact series expansion in terms of the so-called spinning parameter $\mathfrak{s}(\mathbf{n})$. The modal cut-off criterion opens up a way to study the qualitative accuracy of the piston approximation \cite{ICE2, ICE3} numerically. 

As a second application, we investigate the time scale needed for the relaxation of the system in Fig.~\ref{iTD} to the quasi-stationary state, i.e., when exclusively the external pressure $p_{ex}$ governs  the dynamics of the internal pressure $p$ and the membrane displacements $u_{0/L}$. In particular we demonstrate that the quasi-stationary state is sufficient for the description of the system as depicted in Fig.~\ref{iTD}\,(e) in first-order perturbation theory since other contributions will become at least second-order effects in a time scale $T<T_{\rm{neuro}}$ where $T_{\rm{neuro}}\simeq 100\,\text{ms}$ forms the threshold for neuronal information processing and action in animals that use ICE \cite{ICE2, ICE3, ICE1}; say, lizards.

\subsection{Introduction to ABCD context}
\label{introabcd-c}

The ICE model \cite{ICE2, ICE3, ICE1} describes directional hearing in frogs, lizards, crocodilians, and birds by idealizing the eardrums as two sectorial (sectors $\Gamma_0$, $\Gamma_L$ with total opening angle $2\beta$ and radii $a_{\rm{tymp}}$), damped (linear damping coefficient $\alpha$), elastic membranes with displacement field $u_0,u_L$; cf. Fig.~\ref{iTD}\,(e). The membranes respond linearly to the difference in pressures received at both sides of the membranes, $\Psi=p-p_{\rm{ex}}$, which induce the acoustic boundary-condition dynamics, ABCD. The pressure $p$ denotes the acoustic pressure in the cylindrical interaural cavity $\Omega=\Omega(t)$. Both are instrumental in coupling the membranes (radii $a_{\rm{tymp}} < a_{\rm{cyl}}$, interaural distance $L$), with $p_{\rm{ex}}$ as external acoustic signal stimulus. Realistic, overall, geometries are depicted in Fig.~\ref{iTD}\,(a)--(d), whereas Fig.~\ref{iTD}\,(e) is their mathematical abstraction. Details on the biological experiments conducted in lizards that served as a basis for the ICE model can be found elsewhere \cite{Chris1, Chris4, Chris2, Manley2, Manley1, Chris3}. 

In Section~\ref{patag} we have seen that the acoustic wave equation in a cylinder with time-dependent end-caps contained in \{$\Omega(t)$\} can be written in first-order perturbation theory as the deviation from the acoustic wave equation on the stationary cylinder $\Omega_0$ with homogeneous boundary conditions. The boundary conditions of the perturbed problem originate as perturbations from the time-varying end-caps $u_{0,L}(t,\Gamma_{0/L})$ around the stationary end-caps $\lbrace 0/L\rbrace\times\Gamma_{0/L}$.

The analysis of Section \ref{ICE+ABCD} has resulted in a formulation of the ICE equations (\ref{3DW}) \& (\ref{2DW}) in terms of first-order perturbation theory, viz., equations (\ref{sumeq1}) and (\ref{sumeq2}). With the definition of pressure difference $\Psi=p-p_{\rm{ex}}$ driving the membranes, the equations are given by linear second-order partial differential equations on exclusively stationary domains,
\begin{align}\label{toyicemodel}
\begin{split}
&\dfrac{\partial^2 p}{\partial t^2} - c^2\Delta p = \rho_0c^2\dfrac{\partial^2 u_0}{\partial t^2}\, \delta(\mathbf{x}\in\lbrace 0\rbrace\times\Gamma_0) + \rho_0c^2\dfrac{\partial^2 u_L}{\partial t^2}\delta(\mathbf{x}\in\lbrace L\rbrace\times\Gamma_L),\\
&\left(\dfrac{\partial^2 u_0}{\partial t^2} + 2\alpha\dfrac{\partial u_0}{\partial t}\right) - c_m^{2}\Delta_2 u_0 = \dfrac{1}{\rho_m d}\Psi(x=0),\\
&\left(\dfrac{\partial^2 u_L}{\partial t^2} + 2\alpha\dfrac{\partial u_L}{\partial t}\right) - c_m^{2}\Delta_2 u_L = \dfrac{1}{\rho_m d}\Psi(x=L).
\end{split}
\end{align}
As before, $\Delta$ is a three-dimensional Laplacian and $\Delta_2$ refers to the two-dimensional eardrums. We first focus on the underlying physics, then turn to the mathematics proper, and explain the meaning of the different variables below. The pressure $p$ depends on $\mathbf{x}=(x,y,z)\in\Omega_0$ while $u_0$ and $u_L$ are eardrum deflections in the $x$-direction at $x=0$ and $x=L$ depending only on $\mathbf{y}=(y,z)$, as do $\Psi(x=0)$ and $\Psi(x=L)$. Vedurmudi et al. \cite{ICE2} have introduced the so-called \emph{piston approximation} to approximate the above system (\ref{toyicemodel}) of partial differential equations. The idea is this. Since air on a small scale is ``fairly'' incompressible, we replace the time- and space-dependent deviation of the end-caps by their spatial average, viz., $(t,x)\mapsto\int_{\Gamma_{0/L}} dy \, dz \, u_{0/L} (t,y,z,x)$, which effectively results in an air-shifting piston of the same surface, whose position depends on time only. By focusing on the plane-wave mode, the effect of the membranes is one of a system of two piston membranes operating on the cavity $\Omega$. In this section, the decoupling of the equations in first-order perturbation theory is investigated from the point of view of our time-dependent perturbation theory based on Duhamel's principle \cite[\S XIV.5]{Yosida}.

Below we will also introduce the so-called spinning parameter as a quantity to measure how strongly a certain acoustic mode $\mathbf{n}=(n_1,n_2,n_3)$ of the cavity propagates as compared to the associated plane wave mode $(0,0,n_3)$. The spinning parameter will allow us to derive the piston approximation starting from a generalization of the evanescent modal cutoff criterion from duct acoustics \cite{Filippi}, the so-called spinning-mode expansion. Finally, the perturbative scheme to be developed allows an investigation of the relaxation dynamics of the system into the quasi-stationary state, i.e., when the damping of the membrane amplitudes $u_0,u_L$ has reduced them to a level that is negligible in first-order of the coupling strength 
\begin{align}
\label{cpl}
\mathfrak{g}=\rho_0/\rho_m \simeq 10^{-3} \ ,
\end{align}
the ratio between the density $\rho_0$ of air and that of the tympanic membrane, viz., $\rho_m$. As a starting point of our analysis we take (\ref{toyicemodel}), i.e., the ICE equations on stationary domains,
\begin{align}\label{icemodel}
\begin{split}
&\dfrac{\partial^2 p}{\partial t^2} - c^2\Delta p = \rho_0c^2\dfrac{\partial^2 u_0}{\partial t^2}\delta(\mathbf{x}\in\lbrace 0\rbrace\times\Gamma_0) + \rho_0c^2\dfrac{\partial^2 u_L}{\partial t^2}\delta(\mathbf{x}\in\lbrace L\rbrace\times\Gamma_L),\\
&\left(\dfrac{\partial^2 u_0}{\partial t^2} + 2\alpha\dfrac{\partial u_0}{\partial t}\right) - c_m^{2}\Delta_2 u_0 = \dfrac{1}{\rho_m d}\Psi(x=0),\\
&\left(\dfrac{\partial^2 u_L}{\partial t^2} + 2\alpha\dfrac{\partial u_L}{\partial t}\right) - c_m^{2}\Delta_2 u_L = \dfrac{1}{\rho_m d}\Psi(x=L).
\end{split}
\end{align}
The speed of sound in air is denoted by $c$ while $c_m$ is the wave-propagation velocity of flexural waves in the tympanic membranes. Furthermore, $\Delta$ represents the three-dimensional Laplacian in cylindrical coordinates \cite{Polyanin} on $\Omega$, $\Delta_2$ the two-dimensional Laplacian in polar coordinates \cite{Polyanin} on $\Gamma_0/\Gamma_L$ in the equations of motion for $u_0/u_L$. Finally, $\rho_0$ and $\rho_m$ denote the mass density of air and tympanic membranes; see \cite{ICE2} for numerical values. $\Psi$ denotes the driving pressure difference for the eardrum,
\begin{align}\label{blodefunktion}
\Psi(\mathbf{y},x,t) = \left[p(\mathbf{y},x,t) - p_{\rm ex}(t,x)\right]\delta(\mathbf{y}\in\Gamma_0) + \left[p(\mathbf{y},x,t) - p_{\rm ex}(t,x)\right]\delta(\mathbf{y}\in\Gamma_L)
\end{align}
where $\delta(\mathbf{y}\in\Gamma_{0,L})$ denotes the surface-Dirac-delta distribution \cite{surfdelta} for the stationary end-caps $\lbrace 0/L\rbrace\times\Gamma_{0/L}$. 

On the basis of the geometry of the problem \cite{ICE2, ICE3, ICE1}, we take the external sound stimulus (see \cite{Chris1, Chris4, Chris2, Chris3} for experimental realizations) as a plane wave propagating in $x$-direction,
\begin{align}\label{pex}
p_{\rm ex}(x) = \left\lbrace\begin{array}{cc}p_0e^{i\omega t}e^{ikL/2} & :x=0\\
p_0e^{i\omega t}e^{-ikL/2}&: x=L\\
0 & \quad \,\text{elsewhere}\end{array}\right.
\end{align}
Experimental methods for adjusting the angular frequency $\omega$ and the wave-vector's $x$-component $k$ have been discussed in the literature \cite{Chris1, Chris4, Chris2, Chris3}. Mathematically, the initial/boundary-value problem becoming well-defined is achieved by suitably settling initial and boundary conditions \cite{Jackson, Zeidler2}. The tympana are clamped at the boundaries of the sectors $\Gamma_0, \Gamma_L$, \cite{ICE2, ICE3, ICE1}; cf.  Fig.~\ref{iTD}\,(e). This corresponds to homogeneous Di
richlet boundary conditions $u_0(t,\mathbf{y}) = 0\text{ on }\partial\Gamma_0$ and $u_L(t,\mathbf{y}) = 0\text{ on }\partial\Gamma_L$ associated with the two-dimensional Laplacian $\Delta_2$ in the membrane equations on $\Gamma_0$ and $\Gamma_L$. 

Acoustic linearization of Euler's equation \cite{Filippi,Temkin,Williams} results in Neumann boundary conditions for the three-dimensional Laplacian $\Delta$ in the acoustic wave equation on the cylinder $\Omega(t)$. In Section~\ref{patag}, while deriving equation (\ref{dummyeq}), we have shown that we can translate the coupling of the acoustic wave equation and the membrane equations to a source term when we specialize homogeneous boundary conditions in the inhomogeneous wave equation. That is, we set $\partial_{\mathbf{\hat{n}}} p=0\text{ auf }\partial\Omega$ at the cost of including a source term. The associated eigenvalue problem $\Delta\vert\mathbf{n}\rangle = \lambda_{\mathbf{n}}\vert\mathbf{n}\rangle$ admits a solution by means of separation of variables \cite{Polyanin},
\begin{align}\label{nvec}
\vert\mathbf{n}\rangle = \Lambda^{-1}_{\mathbf{n}}J_{n_2}\left(\dfrac{\mu_{n_1,n_2}r}{a_{\rm cyl}}\right)e^{in_2\phi}\cos\left(\dfrac{n_3\pi}{L}x\right) ,\quad \lambda_{\mathbf{n}} = -\left[\dfrac{\mu_{n_1,n_2}}{a_{\rm cyl}^2} + \left(\dfrac{n_3\pi}{L}\right)^2 \right].
\end{align}
Application of separation of variables to the eigenvalue problem for $\Delta_2$ on $\Gamma_0$ and $ \Gamma_L$ for $\Delta_2\vert\mathbf{k}\rangle=\gamma_k\vert\mathbf{k}\rangle$, results in \cite{Polyanin}
\begin{align}\label{kvec}
|\mathbf{k}\rangle = \Lambda^{-1}_{\mathbf{k}}J_{q}\left(\dfrac{\nu_{k_1, k_2}}{a_{\rm tymp}}r\right)\sin\left(\dfrac{k_2\pi\phi}{2\pi - 2\beta}\right) ,\quad\gamma_{\mathbf{k}} = -\dfrac{\nu_{k_1, k_2}^2}{a_{\rm tymp}^2}, \,\, q \equiv \dfrac{k_2\pi}{2(\pi - \beta)}.
\end{align}
In (\ref{nvec}) and (\ref{kvec}) we have used Dirac's bra-ket notation \cite[\S6]{Dirac2}, where 
$\mathbf{n}=(n_1,n_2,n_3)$ $\in\mathbf{N}^3$ and $\mathbf{k}=(k_1,k_2)\in\mathbf{K}^2$ symbolizes the eigenfunctions with discrete label $\mathbf{n}$, $J_q(x)$ is a Bessel function of first kind of order $q$, $\mu_{n_1,n_2}$ is the position of the $n_1$-th extremum of $J_{n_2}$, and $\nu_{k_1,k_2}$ is the $k_1$-th zero of $J_q$ with $q=q(k_2)=k_2\pi(2\pi-2\beta)^{-1}$. 
Furthermore, $\Lambda^{-1}_{\mathbf{k}}$ and $\Lambda^{-1}_{\mathbf{n}}$ are normalization constants \cite{Polyanin}. Since $q\propto \pi k_2$ with rational proportionality constant $(2\pi-2\beta)^{-1}$ \cite{ICE2}, the sets of eigenfunctions $\lbrace\vert\mathbf{n}\rangle\rbrace, \mathbf{n}\in\mathbf{N}^3$ as well as $\lbrace\vert\mathbf{k}\rangle\rbrace, \mathbf{k}\in\mathbf{K}^2$ are complete and orthonormal function systems on $L_2(\Omega)$, $L_2(\Gamma_0)$, and $ L_2(\Gamma_L)$, respectively \cite{Polyanin, Zeidler2}. 

For modeling external sound sources, the membrane system (\ref{icemodel}) must be driven exclusively by $p_{\rm{ex}}$ and $p$. This is a physical interpretation of the mathematical choice of the $0$-th iterate as proposed before. Correspondingly, the initial conditions read
\begin{align}\label{initial}
\begin{split}
p^{(0)}:=p(t=0)=0\quad &,\quad\partial_t p(t=0)=0,\\
u^{(0)}:=u_0(t=0)=0\quad &,\quad\partial_t u_0(t=0)=0,\\
u_L^{(0)}:=u_{L}(t=0)=0\quad &,\quad\partial_t u_L(t=0)=0.
\end{split}
\end{align}

\subsection{Applying Duhamel's principle to ICE}
\label{D+ICE}
The three partial differential equations with partial derivatives up to second order in (\ref{icemodel}) can be transformed into three two-component partial differential equations of first order in the time derivative by introducing the auxiliary fields $q=\partial_t p, v_0 = \partial_t u_0$, and $v_L = \partial_t u_L$. Defining  $u(\mathbf{y},t)=e^{-\alpha t}w(t,\mathbf{y},t)$ reduces the damped membrane equation for $u_0$ and $u_L$ to massive Klein-Gordon equations \cite{Hassani2} for $w_0,w_L,\partial_t w_0$, and $\partial_t w_L$. Duhamel's principle \cite[\S XIV.5]{Yosida} allows expressing the solutions to (\ref{icemodel}) as a convolution integral of an integration kernel depending on the Laplacians $\Delta$, $\Delta_2$, and the source terms, respectively. Carrying out the procedure \cite{Zeidler2} and subsequently reconverting to the physical fields $p, u_0$ and $u_L$ gives to us together with the initial conditions (\ref{initial}), we get the solution
\begin{align}
\label{duhamelICE}
\begin{split}
p(t,\mathbf{x}) &= \rho_0c^2 \int_{0}^{t}d\tau\, \mathfrak{sin}(t-\tau|\Delta)\left[\dfrac{\partial^2 u_0}{\partial \tau^2}(\tau,\mathbf{x}))\mathds{1}_{\Gamma_0}+\dfrac{\partial^2 u_L}{\partial \tau^2}(\tau,\mathbf{x})\mathds{1}_{\Gamma_L}\right],\\
u_0(t,\mathbf{y})&=\dfrac{1}{\rho_m d}\int_{0}^{t}d\tau\, e^{-\alpha(t-\tau)}\mathfrak{sin}(t-\tau|\Delta_2, \alpha) \left[p(\mathbf{y},0,\tau)-p_{\rm{ex}}(\tau,0)\right],\\
u_L(t,\mathbf{y})&=\dfrac{1}{\rho_m d}\int_{0}^{t}d\tau\, e^{-\alpha(t-\tau)} \mathfrak{sin}(t-\tau|\Delta_2, \alpha)\left[p(\mathbf{y},L,\tau)-p_{\rm{ex}}(\tau,L)\right]
\end{split}
\end{align}
where 
\begin{equation}\label{Sin}
\mathfrak{sin}(t|\Delta) = \dfrac{\sin\left(t\sqrt{-c^2\Delta}\right)}{\sqrt{-c^2\Delta}} \ , \quad  \mathfrak{sin}(t|\Delta_2, \alpha) = 
\dfrac{\sin\left(t\sqrt{-(c_m^2\Delta_2 + \alpha^2)}\right)}{\sqrt{-(c_m^2\Delta_2 + \alpha^2)}} \ .
\end{equation}
Here $-\Delta$ and $-\Delta_2$ are essentially self-adjoint, even positive, operators on suitable domains \cite{Yosida} in the Hilbert spaces $L^{2}_{\rm N}(\Omega)$ for $\Delta$ and $L^{2}_D(\Gamma_{0/L})$ for $\Delta_2$ on the eardrums in equilibrium, $\Gamma_0$ and $\Gamma_L$. The indices ``$\rm N$'' and ``$\rm D$'' refer to Neumann and Dirichlet boundary conditions, respectively, assigned to $\Delta$ and $\Delta_2$ on $\Omega \subset \mathbb{R}^3$ and $\partial\Gamma_{0/L}$ for the eardrums, respectively. 

The standard tool to solve wave equations in ducts is Green's functions technology in conjunction with an eigenfunction expansion \cite{Jackson, Williams}. Since the solution to the partial differential equation system is unique, the equivalence between the two approaches can be demonstrated as follows. We define the pressure amplitudes $\mathcal{P}_{\mathbf{n}}(t)$ and membranes amplitudes $\mathcal{U}_{0/L,\mathbf{k}}(t)$ and, using the spectral decomposition \cite{Kato,Zeidler2}, we find
\begin{align}
\label{decomp}
p(t,\mathbf{x}) = \sum_{\mathbf{n}\in\mathbf{N}^3}\mathcal{P}_{\mathbf{n}}(t)\vert\mathbf{n}\rangle\: ,\: u_{0/L}(t,\mathbf{y}) = \sum_{\mathbf{k}\in\mathbf{K}^2}\mathcal{U}_{0/L,\mathbf{n}}(t)\vert\mathbf{k}\rangle\:  .
\end{align}
Taking advantage of (\ref{cpl}), we use  the parameter $\mathfrak{g}=\rho_0/\rho_m \approx 10^{-3}$ as the coupling strength $\mathfrak{g}$ in the above Duhamel representations of $p$, $u_0$, and $u_L$. 
Our parameter $\mathfrak{g}$ is independent of length scales of the membranes in contrast to elsewhere; see, e.g., Kriegsmann, Norris, and Rice \cite{Kriegsmann2,Kriegsmann1}. We use the coupling strength $\mathfrak{g}$ to eliminate one of the mass densities $\rho_0$ and $\rho_m$ in the above Duhamel expansion, and perform an iteration in $\mathfrak{g}$. 

Again by the spectral theorem \cite{Kato,Zeidler2}, the operator sine functions \cite{semiop3, Opsincos} can be expressed \cite{Duffy} in terms of the Green's functions $G_{\mathbf{n}}$ and $H_{\mathbf{k}}$ for the harmonic oscillator and damped harmonic oscillator, respectively. Using the eigenfunctions and eigenvectors in (\ref{nvec}) and (\ref{kvec}) as well as the operator functions $\mathfrak{sin}(t|\Delta)$ and $\mathfrak{sin}(t|\Delta_2, \alpha)$ defined in (\ref{Sin}), we obtain
\begin{align}\label{green}
\begin{split}
G_{\mathbf{n}}(t-\tau)&=\Theta(t-\tau)\langle\mathbf{n}\vert\mathfrak{sin}(t-\tau\vert\Delta)\vert\mathbf{n}\rangle,\\
H_{\mathbf{k}}(t-\tau)&=\Theta(t-\tau)\langle\mathbf{k}\vert e^{-\alpha(t-\tau)}\mathfrak{sin}(t-\tau\vert\Delta_2, \alpha)\vert\mathbf{k}\rangle.
\end{split}
\end{align}
The solution (\ref{duhamelICE}) becomes a solution for the pressure and membrane amplitudes. Furthermore, noting the identity $\mathfrak{g}\,\rho_0^{-1}=\rho_m^{-1}$ in (\ref{duhamelICE}), we obtain expressions for the pressure and membrane amplitudes
\begin{align}
\label{ICE}
\begin{split}
\mathcal{P}_{\mathbf{n}}(t)&=\rho_0c^2\int_{0}^{\infty}d\tau G_{\mathbf{n}}(t-\tau)\partial_{\tau}^2\left\langle\mathbf{n}\vert u_0(\tau,\mathbf{x}))\right\rangle_{\Gamma_0}\\
& +\rho_0c^2\int_{0}^{\infty}d\tau G_{\mathbf{n}}(t-\tau)\partial_{\tau}^2\left\langle\mathbf{n}\vert u_L(\tau,\mathbf{x})\right\rangle_{\Gamma_L},\\
\mathcal{U}_{0,\mathbf{k}}(t)&=\dfrac{\mathfrak{g}}{\rho_0 d}\int_{0}^{\infty}d\tau H_{\mathbf{k}}(t-\tau)\left\langle\mathbf{k}\vert\left[p(r,\phi,0,\tau)-p_{\rm{ex}}(\tau,0)\right]\right\rangle_{\Gamma_0},\\
\mathcal{U}_{L,\mathbf{k}}(t)&=\dfrac{\mathfrak{g}}{\rho_0 d}\int_{0}^{\infty}d\tau H_{\mathbf{k}}(t-\tau)\left\langle\mathbf{k}\vert\left[p(r,\phi,L,\tau)-p_{\rm{ex}}(\tau,L)\right]\right\rangle_{\Gamma_L}.
\end{split}
\end{align}
Comparing with standard solutions such as those of Jackson \cite{Jackson}, we see that the solutions (\ref{ICE}) are precisely those for inhomogeneous wave equations that are obtained by means of Green's functions. In particular, each of $\mathcal{P}_{\mathbf{n}}(t)$ and $\mathcal{U}_{0/L,\mathbf{k}}(t)$ satisfies, irrespectively of the other amplitudes, ordinary differential equations describing harmonic or damped harmonic oscillators \cite{Hassani2, Jackson}. We now apply the Picard-Lindel\"{o}f iteration \cite{Zeidler2} to (\ref{ICE}) using the initial conditions (\ref{initial}) as zeroth iterate. Working in first order in the coupling strength $\mathfrak{g}$, the iteration scheme becomes
\begin{align}
\label{picICE}
\begin{split}
\mathcal{P}^{(l+1)}_{\mathbf{n}}(t)&=\rho_0c^2\int_{0}^{\infty}d\tau G_{\mathbf{n}}(t-\tau)\partial_{\tau}^2\left\langle\mathbf{n}\vert u_0^{(l)}(\tau,\mathbf{x}))\right\rangle_{\Gamma_0}+ \\ &\,\,\,\,\,\,\,\,\rho_0c^2\int_{0}^{\infty}d\tau G_{\mathbf{n}}(t-\tau)\partial_{\tau}^2\left\langle\mathbf{n}\vert u_L^{(l)}(\tau,\mathbf{x})\right\rangle_{\Gamma_L},\\
\mathcal{U}^{(l+1)}_{0,\mathbf{k}}(t)&=\dfrac{\mathfrak{g}}{\rho_0 d}\int_{0}^{\infty}d\tau H_{\mathbf{k}}(t-\tau)\left\langle\mathbf{k}\vert\left[p^{(l)}(r,\phi,0,\tau)-p_{ex}(\tau,0)\right]\right\rangle_{\Gamma_0},\\
\mathcal{U}^{(l+1)}_{L,\mathbf{k}}(t)&=\dfrac{\mathfrak{g}}{\rho_0 d}\int_{0}^{\infty}d\tau H_{\mathbf{k}}(t-\tau)\left\langle\mathbf{k}\vert\left[p^{(l)}(r,\phi,L,\tau)-p_{ex}(\tau,L)\right]\right\rangle_{\Gamma_L}. 
\end{split}
\end{align}
These are the iteration equations to solve the first-order perturbation equations (\ref{sumeq1}) and (\ref{sumeq2}) derived in subsection~\ref{patag}.

\section{Acoustic applications: Piston approximation of the dynamics}
\label{applac}

We now turn to derivation of the piston approximation, which is both a mathematically convenient and a physically appealing way of obtaining explicit solutions. Moreover, we are going to derive a first-oder approximation of the dynamical evolution of the ICE system for all times $t \ge 0$. In so doing we will also see how to proceed in similar situations as they occur, e.g., in acoustics, too often. Finally, we analyze the (in practice fast) approach to the so-called quasi-stationary state as a consequence of damping ($\alpha >0$)

\subsection{Application 1: Derivation of the piston approximation}
\label{apple1}

The first application of the formalism presented in the previous section is the derivation of the piston approximation introduced by Vedurmudi et al. \cite{ICE2}, which is practical in the sense that it produces analytically exact solutions that can be used to match with the experimental findings on directional hearing in e.g. lizards \cite{ICE2}. The external signal is taken to be a pure tone with angular frequency $\omega$, which is switched on at time $t=0$. In view of the linearity of the system, a pure tone is not a restriction of generality. The asymptotic state, which varies harmonically with  angular frequency $\omega$,  is called quasi-stationary. 
We need the perturbative solution to the three-dimensional initial-boundary value problem (\ref{icemodel}) as it approaches the quasi-stationary state for $t\gg\alpha^{-1}$. In first order in $\mathfrak{g}$, the iterative scheme (\ref{picICE}) with zeroth iterates $p^{(0)}, u_{0}^{(0)}$ and $u_L^{(0)}$ as in (\ref{initial}) gives the dynamical evolution for the membrane equations,
\begin{align}
\label{quasimemdis}
u_{0/L}(t,\mathbf{y})= -\dfrac{\mathfrak{g}}{\rho_0 d}e^{i\omega t}\sum_{\mathbf{k}\in\mathbf{K}^2}\dfrac{\vert\mathbf{k}\rangle \langle\mathbf{k}\vert p_{0}e^{(\pm)^{0/L}ikL/2}\rangle_{\Gamma_{0/L}}}{-\omega^2-c^2_m\gamma_{\mathbf{k}} + 2i\alpha\omega} 
\end{align}
where $u_{0/L}$ stands for either $u_{0}$ or $u_{L}$. We can directly get the above result by substituting $p^{(0)}=0$ into the last two equations of (\ref{picICE}), which yields the amplitudes $\mathcal{U}_{0/L,\mathbf{k}}(t)$ in the decomposition (\ref{decomp}), in terms of tractable integrals. The integrals depend on $t$ and can be readily evaluated by insertion of the external pressure $p_{ex}$ as specified in (\ref{pex}). The expression (\ref{quasimemdis}) coincides with the one found before \cite{ICE2}. 
Furthermore, equation (\ref{picICE}) yields for the acoustic pressure $p=p(t,\mathbf{x})$ the solution
\begin{align}\label{spinlesspress}
p = \mathfrak{g} \, \dfrac{(\omega c)^2}{d}\sum_{\mathbf{n}\in\mathbf{N}^3}\sum_{\mathbf{k}\in\mathbf{K}^2}\left[\left(\left\langle\mathbf{n}\vert\mathcal{U}_{0,\mathbf{k}}\mathbf{k}\right\rangle_{\Gamma_0}+\left\langle\mathbf{n}\vert\mathcal{U}_{L,\mathbf{k}}\mathbf{k}\right\rangle_{\Gamma_L}\right)\dfrac{\left(e^{i\omega t}-R_{\mathbf{n}}(t)\right)\vert\mathbf{n}\rangle}{\omega^2 +c^2\lambda_{\mathbf{n}}}\right].
\end{align}
Abusing notation, the time-dependent dynamics of the membrane amplitudes has been singled out as $\mathcal{U}_{0/L}(t)\equiv\mathcal{U}_{0/L}\exp(i\omega t)$. The resonance function
\begin{align*}
R_{\mathbf{n}}(t) = (c\sqrt{-\lambda_{\mathbf{n}}})^{-1}\left[i\omega\sin\left(c\sqrt{-\lambda_{\mathbf{n}}}(t)\right)+c\sqrt{-\lambda_{\mathbf{n}}}\cos\left(c\sqrt{-\lambda_{\mathbf{n}}}(t)\right)\right]
\end{align*}
guarantees the existence of proper limits as $\omega\rightarrow \sqrt{-c^2\lambda_{\mathbf{n}}}$ in (\ref{spinlesspress}). That is,
\begin{align*}
-\infty <\lim_{\omega\to\sqrt{-\lambda_{\mathbf{n}}}}p(t)<\infty .
\end{align*}
The solution for the cavity pressure $p$ in (\ref{spinlesspress}) is the fully time-dependent three-dimensional solution to the acoustic wave equation presented (but not solved) by Vedurmudi et al. \cite{ICE2} in the asymptotic quasi-stationary state including contributions from the resonance function $R_{\mathbf{n}}(t)$. 

In passing we note that we could have replaced the external signal of the form $\exp(i\omega t)$ by any function $\phi(t)$. In the latter case we could proceed exactly as before but we would not have obtained a solution that is explicit as the present one. 

\paragraph{Spinning parameter} As a next step, we define the spinning parameter $\mathfrak{s}(\mathbf{n})$ as the quotient of the frequency-domain propagator of a mode $\mathbf{n}=(n_1,n_2,n_3)$ and the one of a purely axial mode with the same axial mode number $n_3$,
\begin{align}\label{spinparam}
\mathfrak{s}({\mathbf{n}})\equiv\dfrac{-\omega^2 - c^2\lambda_{(0,0,n_3)}}{-\omega^2 - c^2\lambda_{(n_1,n_2,n_3)}},
\end{align}
Expressing the solution (\ref{spinlesspress}) in terms of (\ref{spinparam}), we find
\begin{align}\label{spinpress}
p = -\dfrac{\omega^2\mathfrak{g}c^2}{d} \!\!\!\!\!  \sum_{\mathbf{n}\in\mathbf{N}^3, \mathbf{k}\in\mathbf{K}^2} \!\!\!\!\!\! \mathfrak{s}(\mathbf{n})\left[\left(\left\langle \mathbf{n}\vert\mathcal{U}_{0,\mathbf{k}}\mathbf{k}\right\rangle_{\Gamma_0}+ \left\langle \mathbf{n}\vert\mathcal{U}_{L,\mathbf{k}}\mathbf{k}\right\rangle_{\Gamma_L}\right)\dfrac{\left(e^{i\omega t}-R_{\mathbf{n}}(t)\right)\vert \mathbf{n}\rangle}{-\omega^2 -c^2\lambda_{n}}\right]
\end{align}
The angular frequency $\omega$ of the external pressure signal $p_{\rm{ex}}$ is fixed. The spinning parameter $\mathfrak{s}(\mathbf{n})$ is a monotonously decreasing function of $\mu_{n_1,n_2}$, depending on the modal number $\mathbf{n}=(n_1,n_2,n_3)$ through the $\mu_{n_1,n_2}$. Specifically, one has $\mathfrak{s}(0,0,n_3)=1$ while $\vert\mathfrak{s}(n_1\neq 0,n_2\neq 0, n_3)\vert < 1$. The mode cut-off criterion as used in duct acoustics \cite{Filippi} is obtained from the requirement that the spinning parameter vanishes for all non-axial modes, $\mathfrak{s}_{(n_1,n_2,0)} \stackrel{!}{=}0\Leftrightarrow \omega^2\ll c^2\mu_{n_1,n_2}^2a_{\rm{cyl}}^{-2}.$ In this way, only purely axial modes, i.e., $\mathbf{n} = (0,0,n_3)$ modes, can propagate in the cavity $\Omega$, if $L\gg a_{\rm{cyl}}$ \cite{Filippi,Williams}.

The piston approximation \cite{ICE2} violates the necessary requirement $L\gg a_{\rm cyl}$ for application of the mode cut-off criterion. Expanding (\ref{spinpress}) in $\mathfrak{s}(\mathbf{n})<1$, however, and taking the leading-order contribution in this spinning-mode expansion of (\ref{spinpress}), we nevertheless see that the piston approximation \cite{ICE2} can again be reproduced,
\begin{align}\label{spinpiston}
p \simeq -\dfrac{\omega^2\mathfrak{g}c^2}{d} \, \sum_{n=0}^{\infty}\sum_{\mathbf{k}\in\mathbf{K}^2}\left[\left(\left\langle n\vert\mathcal{U}_{0,\mathbf{k}}\mathbf{k}\right\rangle_{\Gamma_0}+\left\langle n\vert\mathcal{U}_{L,\mathbf{k}}\mathbf{k}\right\rangle_{\Gamma_L}\right)\dfrac{\left(e^{i\omega t}-R_{n}(t)\right)\vert n\rangle}{-\omega^2 -c^2\lambda_{n}}\right].
\end{align}
All acoustic modes with a spinning parameter $\mathfrak{s}(\mathbf{n})<1$ have been neglected as contributions $\mathcal{O}(\mathfrak{s}(\mathbf{n})<1)$ in (\ref{spinpress}). The purely axial modes have been notationally abbreviated by $n$ instead of $(0,0,n_3)$. In the notation of this article, the acoustic pressure becomes in the piston approximation,
\begin{align}
\label{piston}
\begin{split}
p(t,\mathbf{x}) &= -\dfrac{2\omega^2\mathfrak{g}c^2}{dL}\sum_{n=0}^{\infty}\sum_{\mathbf{k}\in\mathbf{K}^2}\left\langle\left\vert\mathcal{U}_{0,\mathbf{k}}\mathbf{k}\right\rangle\right\rangle_{D^2(a_{\rm{cyl}})}\dfrac{\cos\left(\dfrac{n\pi x}{L}\right)}{-\omega^2 - c^2\lambda_n}\left(e^{i\omega t}-R_n(t)\right)\\
&-\dfrac{2\omega^2\mathfrak{g}c^2}{dL}\sum_{n=0}^{\infty}\sum_{\mathbf{k}\in\mathbf{K}^2}\left\langle\left\vert\mathcal{U}_{L,\mathbf{k}}\mathbf{k}\right\rangle\right\rangle_{D^2(a_{\rm{cyl}})}\dfrac{\cos\left(\dfrac{n\pi(L-x)}{L}\right)}{-\omega^2 - c^2\lambda_n}\left(e^{i\omega t}-R_{n}(t)\right) \ .
\end{split}
\end{align}
In the above equation, we have defined the average value
\begin{align*}
\langle f(\mathbf{y}\in\Gamma_{0/L})\rangle_{D^2(a_{\rm cyl})} := \dfrac{1}{\pi a_{\rm cyl}}\int_{D^2(a_{\rm{cyl}})}d^2\mathbf{y}f(\mathbf{y})\mathds{1}_{\Gamma_{0/L}}.
\end{align*}Since the eardrums exhibit vibration amplitudes on the nano-meter scale \cite{Chris1, Chris2}, the above equation describes an approximation of the membranes as flat piston membranes operating on both ends of the cylindrical cavity $\Omega$. It is to be stressed that (\ref{piston}) represents a plane wave propagating only in the axial direction in $\Omega$. The accuracy of the piston approximation can be tested by first considering the spinning parameter and secondly the coupling of the $\vert\mathbf{n}\rangle$-modes to the $\vert\mathbf{k}\rangle$ modes. Normalizing to $a^2_{\rm{cyl}}$, the quantity $a_{\rm{cyl}}^{-2}\mathfrak{s}(\mathbf{n})\langle n_1(r)\vert k_1(r)\rangle_{[0,a_{\rm{tymp}}]}$ is to be assessed numerically. 

The coupling between acoustic and membrane modes is captured in the  $L^2$-inner product in (\ref{spinpress}), 
\begin{align*}\langle n_1(r)\vert k_1(r)\rangle_{[0,a_{\rm{tymp}}]} \equiv\int_{0}^{a_{\rm{tymp}}}dr\, r J_{q}\left(\nu_{k_1,k_2}ra_{\rm{tymp}}^{-1}\right)J_{n_2}\left(\mu_{n_1,n_2}ra_{\rm{cyl}}^{-1}\right).
\end{align*}
The $\phi$-dependent factors in the eigenfunctions (\ref{nvec}) and (\ref{kvec}) contribute multiplicatively as $<1$, i.e., reduce the coupling between different polar modes further. 

\begin{figure}
\label{pistonsim}
\centering
\includegraphics[width = 1\textwidth]{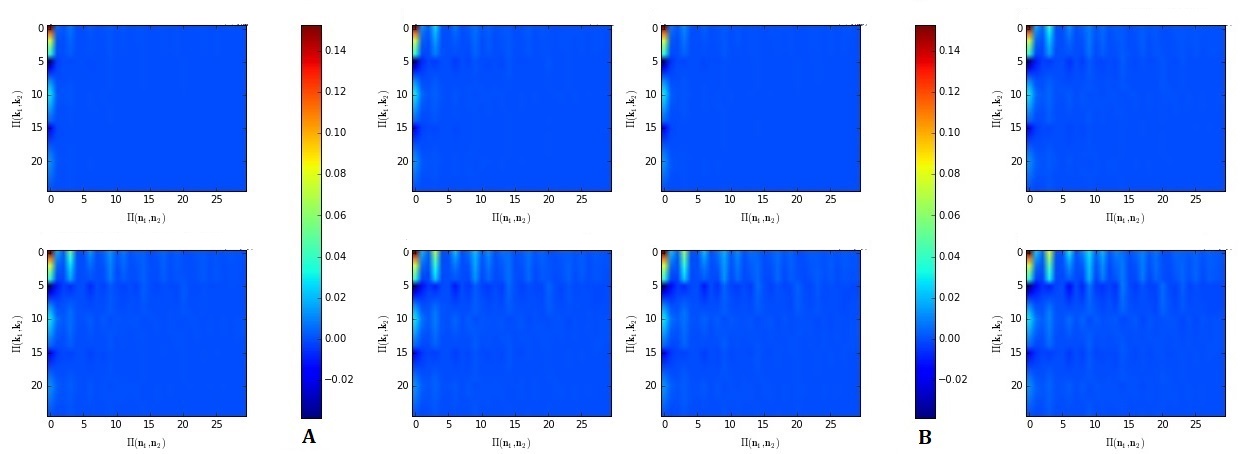}
\caption{Simulation results of $a_{\rm{cyl}}^{-2}\,\mathfrak{s}(\mathbf{n})\langle n_1(r)\vert k_1(r)\rangle_{[0,a_{\rm{tymp}}]}$ for the lizards (a) \emph{Gecko gekko} $(a_{\rm{cyl}}=6.6\,\text{mm}, a_{\rm tymp} = 2.6\,\text{mm}, L = 22\,\text{mm}, \beta = \pi/30)$ at $\omega/(2\pi) = 750\,\text{Hz}$ (left) and (b) \emph{Varanus salvator} $(a_{\rm{cyl}}=6\,\text{ mm}, a_{\rm{tymp}} = 2.6\,\text{ mm}, L = 15.5\,\text{ mm}, \beta = \pi/30)$ at $\omega/(2\pi)=200\,\text{Hz}$. In each plot: left upper corner/right upper corner/left lower corner/right lower corner \& axial mode index $n_3 = 1$/$n_3 = 2$/$n_3 = 3$/$n_3 = 4$. The homogeneous color code ($\approx 0$) everywhere below the main diagonal signals negligible contributions. Only upper-left indicates deviations from the piston approximation.}
\end{figure}

In the practice of the above argument as illustrated by Figure~\ref{pistonsim}, we have mapped the first $30$ acoustic modes at fixed $n_3$  -- using the strict monotony of the extrema $\mu_{n_1,n_2}$ with respect to a suitably ordered index pair -- on the cyclic group of order $30$ ($\mathbb{Z}_{30}$). Likewise, the first $25$ membrane modes are mapped to the cyclic group of order $25$ ($\mathbb{Z}_{25}$), where we use monotony of the zeroes $\nu_{k_1,k_2}$ of the Bessel functions $J_{q(k_2)}$ after re-ordering the index pair. The mapping is formulated as
\begin{align}\label{mappus}
\begin{split}
\Pi:&\mathbb{X}:=\lbrace\mu_{n_1,n_2};n_1,n_2 = 1,2,3,4,5; n_2 = 0\rbrace\rightarrow\mathbb{Z}_{30}\\
&\mu_{n_1,n_2}\mapsto\sharp(\mu_{n'_1,n'_2}\in\mathbb{X}:\mu_{n'_1,n'_2} < \mu_{n_1,n_2})=:\Pi(n_1,n_2)\\
\Pi:&\mathbb{Y}:=\lbrace\nu_{k_1,k_2};k_1,k_2 = 1,2,3,4,5\rbrace\rightarrow\mathbb{Z}_{25}\\
&\nu_{k_1,k_2}\mapsto\sharp(\nu_{k'_1,k'_2}\in\mathbb{X}:\nu_{k'_1,k'_2} < \nu_{k_1,k_2})=:\Pi(k_1,k_2).
\end{split}
\end{align}
The ensuing simulation is shown for the lizard \emph{Gecko gekko} in Figure~\ref{pistonsim}\,(a) at $f = 750\,\text{Hz}$ and \emph{Varanus salvator}, another lizard, in Figure~\ref{pistonsim}\,(b) at $f = 200\,\text{Hz}$ with the parameters as in \cite{ICE2} for the lowest non-trivial axial modes $n_3 = 1,2,3,4$ and $f=\omega/2\pi$ as frequency. The simulated quantity peaks at $\Pi(n_1,n_2)=0$, i.e., becomes maximal at purely axial modes, visible as the dark-red box in the upper left corner of each panel. The axial mode numbers have been chosen to agree with the biological fact that frequencies above $f>10\,\text{kHz}$ are beyond the maximally audible frequency of most, particularly these, lizards. 

Figure~\ref{pistonsim} demonstrates that the piston approximation is qualitatively accurate and quantitatively captures the leading-order contribution (\ref{spinpress}) to the acoustic pressure. The dominant part of each sub-plot in Figure~\ref{pistonsim} is uniformly  blue, which means that the coupling between non-axial acoustic modes and membrane modes is present but in general weak as compared to the tiny square in the upper left corners of each of the four plots for each animal. The upper left-corner is dark red, which means that the coupling between the purely axial acoustic modes and the membrane modes is more dominant than any coupling between spinning acoustic modes and membranes modes in the plotting range. 
\subsection{Application 2: Relaxation of tympanic membranes to their quasi-stationary state}

Because of membrane damping, exemplified by $\alpha >0$ in (\ref{ICE}), the system relaxes to a quasi-stationary state once a pure harmonic has been presented. We expect the relaxation to happen after a time $T_{\rm{eq}}\simeq\alpha^{-1}$ \cite{ICE2, ICE3, ICE1}. The second application of the formalism presented in Section~\ref{appl} therefore is the investigation of the relaxation behavior of the tympanic membranes in the quasi-stationary state. We now show how this proceeds.

The calculations of subsection~\ref{D+ICE} used the assumption $t\gg\alpha^{-1}$ as a working definition of obtaining the quasi-stationary state. Can this definition be refined? Iterating (\ref{picICE}) without assuming $t\gg\alpha^{-1}$ we find that the total membrane displacement $U_{0/L}(t,\mathbf{y})$ is given as a superposition of the membrane amplitudes $u_{0/L}(t,\mathbf{y})$ derived in Section~\ref{gppo} and a correction term including a time-dependent cousin $\mathfrak{h}(t)$ of the coupling strength, specified by
\begin{align}\label{couplstren}
\mathfrak{h}(t)\equiv\mathfrak{g} \exp(-\alpha t)\text{ where }\mathfrak{g}= \rho_0/\rho_m
\end{align}
due to (\ref{cpl}). The quantity $\mathfrak{h}$ is called transient coupling strength. It describes the fluid-structure coupling \cite{Filippi} between the cavity's acoustic wave and the exponentially decaying transient membrane modes from the beginning of the experiment at $t=0$.

\begin{figure}
\label{plotplot}
\centering
\includegraphics[width = 1\textwidth]{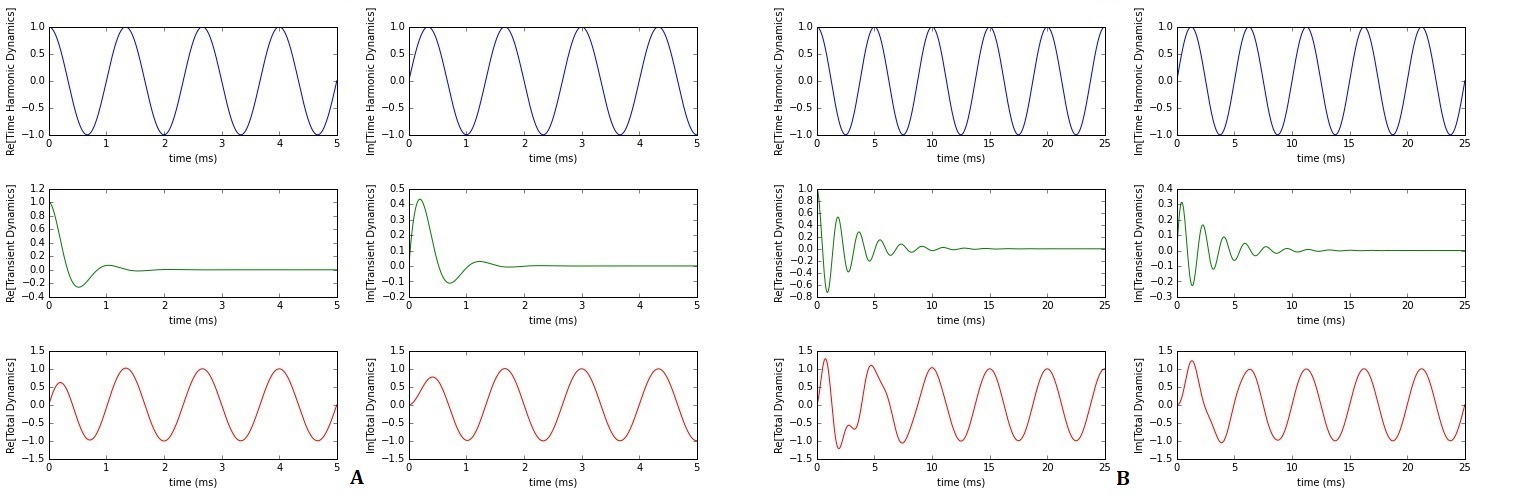}
\caption{(a) Dynamics of the fundamental mode $\vert\mathbf{k}\rangle = \vert 1 1\rangle$ for \emph{Gecko gekko} with $\omega_{11}/(2\pi) = 1050\, \text{Hz}, \alpha = 2611\, \text{Hz})$ during the first $5\,\text{ms}$ for $\omega/(2\pi) = 750\, \text{Hz}$ after exposition to an external sound stimulus. The left column shows the \emph{real} part $(\Re)$ of the dynamics, the right one the \emph{imaginary} part $(\Im)$. The first row depicts the \emph{time-harmonic} dynamics $\exp(i\omega t)$ in blue, the second one the \emph{transient} dynamics  $\exp(-\alpha t) \, t_{\mathbf{k} = (1,1)}(t)$ in green and the third the \emph{total} dynamics $\exp(i\omega t) - \exp(-\alpha t)\, t_{\mathbf{k}=(1,1)}(t)$ in red. (b) Results of the analogous simulation for \emph{Varanus salvator} with $\omega_{11}/(2\pi) = 550\, \text{Hz}, \alpha = 347\, \text{Hz}$ and $\omega/(2\pi)=200\,\text{Hz}$ during the first $25\,\text{ms}$ after exposition to an external sound stimulus. In (a) and (b) the relaxation to the quasi-stationary, or harmonic, asymptotics is fast: after 1 and 5 ms, respectively.}
\end{figure}

Let us define reduced (angular) eigenfrequencies $\omega_{r,\mathbf{k}}\equiv \sqrt{-(c^2_m\gamma_{\mathbf{k}}+\alpha^2)}$ that characterize the eardrum as tympanic membrane. 
Then the complex function describing the relaxation behavior as 
\begin{align}
t_{\mathbf{k}}(t) \equiv (\omega_{r,\mathbf{k}})^{-1}[\omega_{r,\mathbf{k}}\cos(\omega_{r,\mathbf{k}}t) + (\alpha + i\omega)\sin(\omega_{r,\mathbf{k}}t)]
\end{align}
and the transient coupling strength $\mathfrak{h}(t)$ yield through $\mathcal{U}_{0/L,\mathbf{k}}(t) \equiv\mathcal{U}_{0/L,\mathbf{k}}\exp(i\omega t)$, with as usual $0/L$ standing for either $0$ or $L$,  in conjunction with (\ref{picICE}),
\begin{align}
\label{totmemdis}
\begin{split}
U_{0/L}(t,\mathbf{y}) = u_{0/L}(t,\mathbf{y}) + \dfrac{\mathfrak{h}(t)}{\rho_0 d}\sum\nolimits_{\mathbf{k}\in\mathbf{K}^2} t_{\mathbf{k}}(t) \,\mathcal{U}_{0/L,\mathbf{k}}\vert\mathbf{k}\rangle \ .
\end{split}
\end{align}
Figures \ref{plotplot}\,(a) and (b) show a comparison of the time-harmonic quasi-stationary dynamics of the fundamental mode $\vert 1,1\rangle$ of one eardrum with the total dynamics according to (\ref{totmemdis}) for the lizards \emph{Gecko gekko} and \emph{Varanus salvator} \cite{ICE2} during the first $5\,\text{ms}$ and $25$\,ms at sound-stimulus frequency of $750\,\text{Hz}$ and $200\,\text{Hz}$, respectively. The system relaxes for both lizards in a time \cite[Appendix]{ICE3} $T_{\rm{eq}} \approx 3 \,\text{ms} \ll T_{\rm{neuro}}\approx 100\,\text{ms}$ (mostly, $T_{\rm{eq}}< 1$\,ms) to the quasi-stationary state characterized by an almost purely time-harmonic dynamics according to (\ref{quasimemdis}). 

For the calculation of transient corrections $\mathcal{P}_{\mathbf{n}}^{(\rm{tr})}(t)$ to the quasi-stationary pressure amplitudes $\mathcal{P}_{\mathbf{n}}(t)$ according to (\ref{picICE}), we observe the scaling of $\mathcal{P}_{\mathbf{n}}^{(\rm{tr})}(t)$ as $\mathfrak{h}(t)$ as defined in (\ref{couplstren}). Furthermore, the decomposition $\mathcal{P}_{\mathbf{n}}^{(\rm{tot})}(t) \equiv \mathcal{P}_{\mathbf{n}}(t)+\mathcal{P}_{\mathbf{n}}^{(\rm{tr})}(t)$ of the total pressure amplitudes $\mathcal{P}_{\mathbf{n}}^{(\rm{tot})}(t)$ provides a calculational improvement. Converting (\ref{picICE}) into a harmonic oscillator equation for $\mathcal{P}_{\mathbf{n}}^{\rm{(tr)}}(t)$ allows \cite{Hassani2} by defining the quantities $C_{\mathbf{k},\rm{e}} \equiv \alpha^2 - \omega_{r,\mathbf{k}}^2 - 2\alpha(\alpha + i\omega)$, $C_{\mathbf{k},\rm{o}} \equiv \omega_{r,\mathbf{k}}^{-1}\left((\alpha^2 - \omega_{r,\mathbf{k}}^2)(\alpha + i\omega) + 2\alpha\omega_{r,\mathbf{k}}^2\right)$ and $k_{\mathbf{n},\mathbf{k}} \equiv \left((-\alpha + i\omega_{r,\mathbf{k}})^2 + c^2\lambda_{n}\right)^{-1}$ the solution
\begin{align*}
\begin{split}
\mathcal{P}^{(\rm{tr})}_{\mathbf{n}}(t)=& -\dfrac{\mathfrak{h}(t)c^2}{d}\sum_{\mathbf{k}\in\mathbf{K}^2}\left[C_{\mathbf{k},\rm{e}}\Re\left(k_{\mathbf{n},\mathbf{k}}e^{i\omega_{r,\mathbf{k}}t}\right) + C_{\mathbf{k},\rm{o}}\Im\left(k_{\mathbf{n},\mathbf{k}}e^{i\omega_{r,\mathbf{k}}t}\right)\right]\left\langle\mathbf{n}\vert\mathcal{U}_{0,\mathbf{k}}\mathbf{k}\right\rangle_{\Gamma_0}\\
&-\dfrac{\mathfrak{h}(t)c^2}{d}\sum_{\mathbf{k}\in\mathbf{K}^2}  \left[C_{\mathbf{k},\rm{e}}\Re\left(k_{\mathbf{n},\mathbf{k}}e^{i\omega_{r,\mathbf{k}}t}\right) + C_{\mathbf{k},\rm{o}}\Im\left(k_{\mathbf{n},\mathbf{k}}e^{i\omega_{r,\mathbf{k}}t}\right)\right]\left\langle\mathbf{n}\vert\mathcal{U}_{L,\mathbf{k}}\mathbf{k}\right\rangle_{\Gamma_L}
\end{split}
\end{align*}
where $\Re$ and $\Im$ denote real and imaginary part, respectively. Thus both the transient corrections to the membrane and pressure amplitudes scale as $\mathfrak{h}(t)$. Requiring consistency with first-order perturbation theory in $\mathfrak{g}$ ($\mathfrak{h}(t)<\mathfrak{g}^2$) yields a characteristic time scale for the relaxation of the tympanic membranes to the quasi-stationary state ($\mathfrak{g}\simeq 10^{-3}$ and typically \cite[Appendix]{ICE3} $\alpha\simeq 2\text{-}3 \times 10^3\,\text{Hz}$),
\begin{align}\label{relax}
T_{\rm{eq}} = -\alpha^{-1}\ln\mathfrak{g} \simeq 3 \text{ ms}.
\end{align}
This is far below the threshold time-scale $T_{\rm{neuro}}$, in that $T_{\rm{eq}}\ll T_{\rm{neuro}}\approx 100\,\text{ms}$ above which information processing on the neuronal level is predominantly supported.

\section{Discussion and algorithmic summary}
\label{discas}

It is time to summarize our results and the algorithmic steps that have led to them. This is dome in two different ways. First, we simply formulate, as \emph{fait accomplis},  a theorem based on the geometry of Fig.~\ref{iTD}\,(e). Second, by carefully spelling out the perturbation-theoretic algorithm itself. The geometry of Fig.~\ref{iTD}\,(e) gives rise to eigenfunctions of the unperturbed two- and three-dimensional Laplacians that can be expressed analytically so that smooth, even analytic, extension outside the original domain is straightforward. See Jost \cite[Sections 11.3--11.5]{jost} for the standard regularity results on eigenfunctions; in particular Theorem. 11.3.1 and Corollary 11.4.1.

\subsection{Formulating a theorem}

The theorem below, whose ingredients we have already met on our way to this section, is based on the simple geometry of Fig.~\ref{iTD}\,(e) in conjunction with a few simplifications that we now list \footnote{Though implicit, a carefully spelled-out proof can be obtained from the authors.}. 

\paragraph{Simplifications}
(i)  \emph{Smallness of perturbations}  The perturbations are supposed to be small in the sense that $\max_{(t,x,y,z)}\vert U/L\vert = \epsilon \ll 1$. Thus, $U/L \sim \mathcal{O}(\epsilon)$ where e.g. in acoustics $\epsilon = 10^{-6}$ at most. \hfill\newline
(ii)  \emph{Membrane-like displacements} The modeling assumption is that $\Vert \partial_z^2 U\Vert\ll \Vert(\Delta - \partial_z^2)U\Vert$ so that the equation for $U$ becomes effectively a membrane equation. In other words, the equation axial part of the Laplacian is discarded from the equation for $U$ and we model the two membranes at the $z=0$ and $z=L$ end cap through one equation.\hfill\newline
(iii) \emph{Smallness of perturbations variations}  We assume small variations in the sense that $\partial_z U(t,x,y,z) \ll 1$ so that $U$ can be treated as effectively $z$-independent. This corresponds to $k\cos\theta\ll 2/L$. The assumption is e.g. acoustically realistic \cite{ICE3} in that the source term model as described above becomes an inadequate description if $\theta = 0$, i.e., when a sound source is located along the cylinder's symmetry axis. Then, only the closer of the two, the ipsilateral membrane, is excited by the stimulus. The contralateral membrane resides in the acoustic shadow region and is not affected by the sound stimulus at all. However, if $\theta$ is in the vicinity of $\pi/2$, the cosine approaches $0$ as $\sim -(\theta - \pi/2)$ and the above smallness requirement is satisfied.\hfill\newline
(iv) \emph{Regularity of $p$ in the transverse coordinate} We assume $p$ to be smooth or even analytic -- think of the Bessel functions in the case of Fig.~\ref{iTD}\,(e) -- in the third spatial variable $z$. That is, in the coordinate that is perturbed. We require that $\exp(\epsilon \partial_{\hat{z}}) \mathsf{D}^2 p(t,x,y,z)$ is square integrable over $(0,T)\times\Omega$ for all $T>0$ where $\mathsf{D}^2\in\lbrace\partial_t^2,\partial_x^2,\partial_y^2,\partial_z^2\rbrace$ and $\hat{z} = z/L$ while $\exp(\epsilon \partial_{\hat{z}})$ is the associated translation operator.

\begin{mythm}
\begin{enumerate}
\item\emph{Reducibility via linearization}
Up to an error of order $\epsilon$ to the contribution of the wave operator $\partial_t^2-c^2\Delta$, the wave equation (\ref{3DW}) equals $\partial_t^2 p(t,x,y,z) - c^2\Delta p(t,x,y,z)=0$ on $\Omega_0$. The boundary conditions are pulled back accordingly and turn into $\partial_z p(t,x,y,z=0)=-\rho_0\partial_t^2 u_0(t,x,y)$ and $\partial_z p(t,x,y,z=0) = -\rho_0\partial_t^2 u_L(t,x,y)$, respectively.\hfill\newline
\item\emph{Existence and uniqueness of solutions to the linearized problem}  With homogeneous initial conditions (e.g., $\equiv 0$), the linearized problem on the stationary domain $\Omega$ has one and only one solution, viz., $(p,u_0,u_L)\in C^2(\mathbb{R}^+_0\to H^1(\Omega))\times C^2(\mathbb{R}^+_0\to H^1_0(\Gamma_0))\times C^2(R^+_0\to H^1_0(\Gamma_L))$, if $p$ satisfies homogeneous Neumann boundary conditions on $\partial\Omega$ and $u_0$ and $u_L$  satisfy homogeneous Dirichlet boundary conditions on $\partial\Gamma_0$ and $\partial\Gamma_L$, respectively. In particular, if $\mathfrak{g}\equiv\rho_0/\rho_m\ll 1$, which holds for natural samples, the first-order solution triplet can be found by truncation of Picard-iteration once $\alpha > 0$ in $\mathfrak{g}$.\hfill\newline
\item\emph{Spinning-mode expansion} Once $L > a_{\rm cyl}> 0$ while $p_{0,\rm ex}=p_{0,\rm ex}(t)$ and $p_{L,\rm ex}=p_{L,\rm ex}(t)$, the maximum contribution to the eigenfunction expansion is realized by the purely axial modes. Maximum is to be understood in the sense that the expansion coefficient has maximum modulus.
\item\emph{Relaxation dynamics} In lowest non-trivial order, the transient contribution to the solution of the linearized (in $\epsilon)$ and decoupled (in $\mathfrak{g}$) system of partial differential equations 
becomes negligible for $t\geq -\alpha^{-1}\ln(\mathfrak{g})$. Here $\alpha$ is the linear damping coefficient in the membrane equation (\ref{2DW}) and $\mathfrak{g}$ is the coupling strength stemming from (\ref{cpl}). In other words, after $t\geq -\alpha^{-1} \ln(\mathfrak{g})$ the linearized and decoupled system of wave equations having homogeneous initial conditions turns, up to error contributions scaling as $\mathfrak{g}^2$, indistinguishable from the corresponding quasi-stationary state Helmholtz equations, which have no initial conditions altogether. 
\end{enumerate}
\end{mythm}

By means of the time-dependent, Dirac-like \cite{Dirac1,Dirac2}, perturbation scheme derived in Section~\ref{D+ICE}, in particular, Eq.~(\ref{picICE}), the general ideas presented in Section~\ref{gppo} have been applied, for the sake of definiteness, to the special case of the ICE model \cite{ICE1,ICE2, ICE3}. Through the introduction of the coupling strength $\mathfrak{g}=\rho_0/\rho_m$ as parameter, we could use the iteration scheme (\ref{picICE}) to systematically decouple the previously coupled partial-differential-equation dynamics (\ref{toyicemodel}). In doing so and in conjunction with all operator semigroups being contractive, we have obtained the solution to the three-dimensional wave equation (\ref{spinlesspress}) with acoustically varying boundary conditions in first order of the coupling strength $\mathfrak{g}$; see in particular  Eqs. (\ref{picICE}), (\ref{quasimemdis}) and (\ref{spinlesspress}). Moreover, we can do so in a well-controlled manner, with $0 < \mathfrak{g} \ll 1$ telling us that the first-order approximation is already a good one. 

The method that has been developed here is quite general and can handle any finite cavity in contact with the external auditory world through bounding manifolds \cite{david2}, modulo the extension of the eigenfunctions of the three-dimensional Laplacian for the unperturbed domain, whose two-dimensional boundary is supposed to be smooth. 

Introducing the spinning parameter (\ref{spinparam}) as the quotient of plane-wave axial mode versus Fourier-Bessel axial mode propagation, we have extended the mode cut-off criterion as used in duct acoustics \cite{Filippi}  to a general series expansion. The conventional criterion is reproduced in the limit of vanishing spinning parameter for non-axial modes. By means of the series expansion (\ref{spinpress}), the piston approximation \cite{ICE2} can be justified and the plane-wave result (\ref{spinpiston}) \& (\ref{piston}) can be deduced from the full solution (\ref{icemodel}) to the three-dimensional dynamics. As shown in Section~\ref{introabcd-c}, the piston approximation originates from a truncation of the spinning-mode series expansion (\ref{spinpress}), which is general enough to be transferred to other cylindrical-coordinate models. 

Furthermore, the present formalism allows a consistent perturbative formulation to justify the widely used `quasi-stationarity assumption,' practiced in the present context by Vedurmudi et al. \cite{ICE2}. As (\ref{relax}) shows explicitly, the system quickly relaxes into the quasi-stationary state, the speed of relaxation of course depending on the damping parameter $\alpha$. The relaxation dynamics can now be derived straightforwardly by using linearity of the partial differential operators in a systematic iteration scheme such as (\ref{picICE}). We can write the general membrane and pressure amplitudes as a superposition of transient amplitudes decaying due to the membrane damping and a contribution that carries the time-harmonic (or whatever) dynamics of the external stimulus. Investigating the system for a very short time after external stimulation onset (e.g., $\sim 7\,\text{ms}$), we have seen that the transient contributions are only of sub-leading order and the animal predominantly experiences the external sound stimulus. The Duhamel representation is the key to efficiently handling the transients as well, which as such singles out the present approach uniquely. 

Finally, geometric perturbation theory exemplified by acoustic boundary condition dynamics (ABCD) is based on the fundamental idea that perturbation theory as applied to fluctuating acoustic boundary conditions ought to exploit and, hence, critically depends on the \emph{geometry} of the underlying dynamics. The deviations from equilibrium being quite small, which is a reformulation of the smallness of the coupling strength $\mathfrak{g} \ll 1$, guarantees that not only ABCD gives rise to a valid perturbation scheme but also that the algorithmic structure of geometric perturbation theory is far more general than ABCD and, actually, universal. It inherently unifies the geometry of the perturbations with the dynamically varying boundary conditions. The present arguments are the nucleus of a far more general but also far more involved theory of geometric perturbations, for which we refer to elsewhere \cite{david2}. The universality of geometric perturbation theory has been demonstrated there. 

\subsection{Algorithmic summary} 

Since the aim of the present paper is providing a mathematical basis for the applicability of geometric perturbation theory, we finish it by an algorithmic overview, divided up into seven concrete steps. 

\paragraph{1. General setup} As a first step, we need to identify the acoustic enclosures $(\Omega_i)_{i\in I}$, which may be of any finite size, and the connective vibrational structures $(\Gamma_j)_{j\in J}$ where both $I$ and $J$ are finite sets; cf. Section~\ref{patag}. The domains $(\Omega_i)_{i\in I}$ and $(\Gamma_j)_j$ are assumed to be at least $C^2$-regular, compact, and simply-connected with closed boundary $\partial\Gamma_j$.  In vibro-acoustic models specification is straightforward. 

\paragraph{2. Smallness of perturbations} Next, we need to ensure that the elastic constituents of the general setup only give rise to small perturbations; cf. Sections \ref{patag} and \ref{introabcd-c}.  Smallness is understood in the sense that if $L\equiv\min_{i\in I}\lbrace\sqrt[3]{\text{Vol}(\Omega_i)}\rbrace$ denotes a characteristic length scale for the acoustic enclosures, we have $U/L\ll 1$ where $U$ denotes a characteristic, maximal, length scale for the perturbations $u_j$ of the $(\Gamma_j)_j$ from the equilibrium positions $u_j  \equiv 0$. The perturbations are the displacement fields describing the normal vibration of $\Gamma_j(t)$, i.e., $u_j(t,\Gamma_j)$ for $j\in J$. 

\paragraph{3. Model equations} We specify the dynamical equations such as the wave equations in acoustics; cf. Section~\ref{introabcd-c}. That is, for all $i\in I$ we have an acoustic wave equation on the domain $\Omega_i$ with homogeneous initial conditions $p_i(t=0)=0=\partial_t p_i(t=0)=0$.  Moreover, we have no-slip and, hence, Neumann boundary conditions \cite[Chapter 2]{Temkin} from Euler's equation, $-\rho_0\partial_t v_{n}=\partial_n p$ at the boundaries $\partial\Omega_i$ and we equate the acoustic normal velocity $v_n$ to the displacement $(\pm)_{j,i,k}u_j$ for all $j\in J$ such that $\Gamma_j\subseteq\partial\Omega_i$; cf. (\ref{struc3}). 

\paragraph{4. Unperturbed problem} We derive the discretely labeled eigenfunctions $\lbrace\lbrace\Psi_{\mathbf{n},i}\rbrace_{\mathbf{n}}\rbrace_{i}$ and $\lbrace\lbrace\Phi_{\mathbf{k},j}\rbrace_{\mathbf{k}}\rbrace_{j}$ with $i\in I$ and $j\in J$ as well as their  corresponding eigenvalues -- see Sections \ref{patag} and \ref{D+ICE} -- for the wave operators $\Delta_i$ and $\mathsf{L}_j$ while using homogeneous Neumann and Dirichlet boundary conditions for $\Delta_i$ and $\mathsf{L}_j$, respectively.

\paragraph{5. Systematic perturbation theory} For $p_{\rm{ex}}=0$, the unperturbed solution  is $p^{(0)}_i=0$ and $u^{(0)}_j=0$. We now restore the original $p_{\rm{ex}}\neq 0$ and apply Banach's fixed-point theorem in conjunction with the Duhamel principle \cite[\S XIV.5]{Yosida} and Picard-style iterations so as to obtain a perturbation expansion in terms of the small parameter $\epsilon = \vert U/L\vert$; cf. the example of Section~\ref{D+ICE}. In so doing we can take advantage of the (complete and orthonormal) set of eigenfunctions from the previous step to implement the unperturbed semigroups that occur. 

\paragraph{6. Truncating the perturbation series} We truncate the iterations of the dynamical equations, such as (\ref{picICE}) in subsection~\ref{D+ICE}. We can do so on the basis of the a-priori estimates for the convergence of the Banach fixed-point iteration. In acoustics, a first-order term will do since $\epsilon$ is of the order of $10^{-6}$; except for the disco, which does not occur in nature. In practice, $p_{\rm{ex}}$ is smooth. For details, see Section~\ref{applac}. 

\paragraph{7. Full solution through iteration} One can use the spectral decomposition \cite{Zeidler2} to advantage so as to set up an eigenfunction expansion for the perturbed problem by ``inverting'' the projection from the above step 5. This yields a truncation of the full solution to a first-order problem. For practical applications such as to engineering, capturing the first-order perturbation terms suffices; cf. Eqs. (\ref{quasimemdis}), (\ref{spinlesspress}), and (\ref{piston}). The nonlinear effects are more elaborate and for acoustic applications less rewarding to model, but doable \cite{david2}. 

In the case of axial symmetries for $(\Omega_i)_{i \in I}$, e.g., for cylindrical enclosures, a spinning-mode series expansion can help to cut off spinning modes, although the conventional cut-off criterion \cite[Section 7.2.3, specifically pp. 220--221]{Filippi} is not applicable. The result of the procedure is the piston approximation; see subsection~\ref{apple1}. The spinning-mode series expansion requires a partial separability structure $\Delta_i=\Delta_{\partial\Omega_i}+\partial_z^2$ in the Laplacians $\Delta_i$ on the domains $(\Omega_i)_{i \in I}$. Details on accuracy as well as geometric interpretation of the piston approximation have been outsourced \cite{david2}. Numerically, the spinning-mode series expansion can be implemented easily. 

Finally, geometric perturbation theory intertwines fluctuating geometries and the dynamics that drives them and, in doing so, gives rise to a flexible formalism to analytically treat rather intricate phenomena; for instance, but not only, in acoustics. 

\paragraph{Acknowledgments} The authors thank Prof. Folkmar Bornemann (TUM) for his constructive criticism and Dr.~Anupam Vedurmudi for practical advice regarding the biological physics of audition in ``icy'' animals. 

\bibliography{Heider+vanHemmen_Physica_D_6_litt}     

\begin{thebibliography}{46}
\providecommand{\natexlab}[1]{#1}
\providecommand{\url}[1]{\texttt{#1}}
\expandafter\ifx\csname urlstyle\endcsname\relax
  \providecommand{\doi}[1]{doi: #1}\else
  \providecommand{\doi}{doi: \begingroup \urlstyle{rm}\Url}\fi

\bibitem[Beale(1976)]{Beale2}
J.~Beale.
\newblock Spectral properties of an acoustic boundary condition.
\newblock \emph{Indiana Univ. Math. J.}, 25\penalty0 (9):\penalty0 895--917,
  1976.

\bibitem[Beale and Rosencrans(1974)]{Beale1}
J.~Beale and S.~Rosencrans.
\newblock Acoustic boundary conditions.
\newblock \emph{Bull. Am. Math. Soc.}, 80\penalty0 (6):\penalty0 1276--1278,
  1974.

\bibitem[Christensen-Dalsgaard(2005)]{jcd1}
J.~Christensen-Dalsgaard.
\newblock Directional hearing in non-mammalian tetrapods.
\newblock In A.~N. Popper and R.~R. Fay, editors, \emph{Sound Source
  Localization}, Springer Handbook in Auditory Research, pages 67--123.
  Springer, New York, 2005.

\bibitem[Christensen-Dalsgaard and Carr(2005)]{Chris1}
J.~Christensen-Dalsgaard and C.~E. Carr.
\newblock Directionality of the lizard's ear.
\newblock \emph{J. Exp. Biol.}, 208:\penalty0 1209--1217, 2005.

\bibitem[Christensen-Dalsgaard and Carr(2008)]{Chris4}
J.~Christensen-Dalsgaard and C.~E. Carr.
\newblock Evolution of a sensory novelty: tympanic ears and the associated
  neural processing.
\newblock \emph{Brain. Res. Bull.}, 75\penalty0 (2-4):\penalty0 365--370, 2008.

\bibitem[Christensen-Dalsgaard and Manley(2008)]{Chris2}
J.~Christensen-Dalsgaard and G.~Manley.
\newblock Acoustical coupling of lizard eardrums.
\newblock \emph{J. Assoc. Res. Otol.}, 9\penalty0 (4):\penalty0 407--416, 2008.

\bibitem[Christensen-Dalsgaard et~al.(2011)Christensen-Dalsgaard, Tang, and
  Carr]{cecytjcd}
J.~Christensen-Dalsgaard, Y.~Tang, and C.~E. Carr.
\newblock Binaural processing by the gecko auditory periphery.
\newblock \emph{J. Neurophysiol.}, 105\penalty0 (9):\penalty0 1992--2004, 2011.

\bibitem[Deng et~al.(2006)Deng, Douglas, Kako, Masabumi, and Ichiro]{DengLi}
L.~Deng, C.~C. Douglas, T.~Kako, S.~Masabumi, and H.~Ichiro.
\newblock A novel perturbation expansion method for coupled system of acoustics
  and structures.
\newblock \emph{Comp. Math. Appl.}, 51:\penalty0 1689--1704, 2006.

\bibitem[Dirac(1927)]{Dirac1}
P.~A.~M. Dirac.
\newblock The quantum theory of emission and absorption of radiation.
\newblock \emph{Proc. R. Soc. (London) A}, 114:\penalty0 243--265, 1927.

\bibitem[Dirac(1958)]{Dirac2}
P.~A.~M. Dirac.
\newblock \emph{Quantum Mechanics}.
\newblock Oxford University Press, Oxford, 4th edition, 1958.

\bibitem[Duffy(2015)]{Duffy}
D.~Duffy.
\newblock \emph{Green's Functions with Applications}.
\newblock CRC Press, Boca Raton , FL, 2015.

\bibitem[Evans(2010)]{Evans}
L.~C. Evans.
\newblock \emph{Partial Differential Equations}.
\newblock American Mathematical Society, Providence , RI, 2nd edition, 2010.

\bibitem[Filippi et~al.(1999)Filippi, Habault, and Lefebvre]{Filippi}
P.~Filippi, D.~Habault, and J.~P. Lefebvre.
\newblock \emph{Acoustics: Basic Physics, Theory and Methods}.
\newblock Academic, London, 1999.

\bibitem[Fr\"{o}hlich(1938)]{Frohlich}
H.~Fr\"{o}hlich.
\newblock A solution of the {S}chr\"{o}dinger equation by a perturbation of the
  boundary conditions.
\newblock \emph{Phys. Rev.}, 54:\penalty0 945--947, 1938.

\bibitem[Frota et~al.(2011)Frota, Medeiros, and Vicente]{Beale3}
C.~L. Frota, L.~A. Medeiros, and A.~Vicente.
\newblock Wave equation in domains with non-locally reacting boundary.
\newblock \emph{Differential and Integral Equations}, 24\penalty0
  (11-12):\penalty0 1001--1020, 2011.

\bibitem[Gal et~al.(2003)Gal, Goldstein, and Goldstein]{semiop2}
C.~Gal, G.~Goldstein, and J.~Goldstein.
\newblock Oscillatory boundary conditions for acoustic wave equations.
\newblock \emph{J. Evol. Equ.}, 3:\penalty0 623--635, 2003.

\bibitem[Hassani(2013)]{Hassani2}
S.~Hassani.
\newblock \emph{Mathematical Physics: An Introduction to its Foundations}.
\newblock Springer, New York, 2013.

\bibitem[Heider and van Hemmen(2017)]{david2}
D.~Heider and J.~L. van Hemmen.
\newblock Geometric perturbation theory for a class of fiber bundles, 2017.
\newblock {arXiv:1801.00360 [math-ph]}.

\bibitem[Jackson(1999)]{Jackson}
J.~D. Jackson.
\newblock \emph{Classical Electrodynamics}.
\newblock Wiley Interscience, New York, 3rd edition, 1999.

\bibitem[Jost(2013)]{jost}
J.~Jost.
\newblock \emph{Partial Differential Equations}.
\newblock Springer, New York, 3rd edition, 2013.

\bibitem[Kato(1966)]{Kato}
T.~Kato.
\newblock \emph{Perturbation Theory for Linear Operators}.
\newblock Springer, New York, 1966.

\bibitem[Kriegsmann et~al.(1984)Kriegsmann, Norris, and Reiss]{Kriegsmann2}
G.~Kriegsmann, A.~Norris, and E.~Reiss.
\newblock Acoustic scattering by baffled membranes.
\newblock \emph{J. Acoust. Soc. Am.}, 75\penalty0 (3):\penalty0 685--694, 1984.

\bibitem[Kriegsmann et~al.(1986)Kriegsmann, Norris, and Reiss]{Kriegsmann1}
G.~Kriegsmann, A.~Norris, and E.~Reiss.
\newblock Acoustic pulse scattering by baffled membranes.
\newblock \emph{J. Acoust. Soc. Am.}, 79\penalty0 (1):\penalty0 1--8, 1986.

\bibitem[Lange(2012)]{surfdelta}
R.-J. Lange.
\newblock Potential theory, path integrals and the {L}aplacian of the
  indicator.
\newblock \emph{Journal of High Energy Physics = JHEP}, 11/032:\penalty0 1--46,
  2012.

\bibitem[Manley(1972{\natexlab{a}})]{Manley1}
G.~Manley.
\newblock The middle ear of the {T}okay gecko.
\newblock \emph{J. Comp. Physiol.}, 81\penalty0 (3):\penalty0 239--250,
  1972{\natexlab{a}}.

\bibitem[Manley(1972{\natexlab{b}})]{Manley2}
G.~Manley.
\newblock Frequency response of the middle ear of geckos.
\newblock \emph{J. Comp. Physiol.}, 81\penalty0 (3):\penalty0 251--258,
  1972{\natexlab{b}}.

\bibitem[Pan and Bies(1990{\natexlab{a}})]{Pan1}
J.~Pan and D.~Bies.
\newblock The effect of fluid-structural coupling on sound waves in an
  enclosure - theoretical part.
\newblock \emph{J. Acoust. Soc. Am.}, 87\penalty0 (2):\penalty0 691--707,
  1990{\natexlab{a}}.

\bibitem[Pan and Bies(1990{\natexlab{b}})]{Pan2}
J.~Pan and D.~Bies.
\newblock The effect of fluid-structural coupling on sound waves in an
  enclosure - experimental part.
\newblock \emph{J. Acoust. Soc. Am.}, 87\penalty0 (2):\penalty0 708--717,
  1990{\natexlab{b}}.

\bibitem[Phillips(1952)]{semiop3}
R.~S. Phillips.
\newblock Perturbation theory for semi-groups of linear operators.
\newblock \emph{Trans. Am. Math. Soc.}, 1952:\penalty0 199--221, 1952.

\bibitem[Polyanin(2002)]{Polyanin}
A.~Polyanin.
\newblock \emph{Handbook of Linear Partial Differential equations for
  Scientists and Engineers}.
\newblock CRC Press, Boca Raton, FL, 2002.

\bibitem[Reed and Simon(1972)]{Reed1}
M.~Reed and B.~Simon.
\newblock \emph{Methods of Modern Mathematical Physics {I}: Functional
  Analysis}.
\newblock Academic, San Diego, CA, 1972.

\bibitem[Reed and Simon(1975)]{Reed2}
M.~Reed and B.~Simon.
\newblock \emph{Methods of Modern Mathematical Physics {II}: Fourier Analysis,
  Self-Adjointness}.
\newblock Academic, San Diego, CA, 1975.

\bibitem[Reed and Simon(1978)]{Reed4}
M.~Reed and B.~Simon.
\newblock \emph{Methods of Modern Mathematical Physics {IV}: Analysis of
  Operators}.
\newblock Academic, San Diego, CA, 1978.

\bibitem[Schwartz(1966)]{schwartz}
L.~Schwartz.
\newblock \emph{Th\'{e}orie des {D}istributions}.
\newblock Herman, Paris, 1966.

\bibitem[Shearer and Levy(2015)]{sh+l}
M.~Shearer and R.~Levy.
\newblock \emph{Partial Differential Equations: An Introduction to Theory and
  Applications}.
\newblock Princeton University Press, Princeton, 2015.

\bibitem[Temkin(1981)]{Temkin}
S.~Temkin.
\newblock \emph{Elements of Acoustics}.
\newblock Wiley, New York, 1981.

\bibitem[van Hemmen et~al.(2016)van Hemmen, Christensen-Dalsgaard, Carr, and
  Narins]{ice-editorial}
J.~L. van Hemmen, J.~C.~D. Christensen-Dalsgaard, C.~E. Carr, and P.~M. Narins.
\newblock Animals and ice: meaning, origin, and diversity.
\newblock \emph{Biol. Cybern.}, 110:\penalty0 237--246, 2016.

\bibitem[Vasil'ev and Piskarev(2004)]{Opsincos2}
V.~V. Vasil'ev and S.~I. Piskarev.
\newblock Differential equations in {B}anach spaces {II}: Theory of cosine
  operator functions.
\newblock \emph{J. Sov. Math.}, 122\penalty0 (2):\penalty0 3055--3174, 2004.

\bibitem[Vasil'ev et~al.(1991)Vasil'ev, Krein, and Piskarev]{Opsincos}
V.~V. Vasil'ev, S.~Krein, and S.~I. Piskarev.
\newblock Operator semigroups, cosine operator functions, and linear
  differential equations.
\newblock \emph{J. Sov. Math.}, 54\penalty0 (4):\penalty0 1042--1129, 1991.

\bibitem[Vedurmudi et~al.(2016{\natexlab{a}})Vedurmudi, Goulet,
  Christensen-Dalsgaard, Young, Williams, and van Hemmen]{ICE2}
A.~P. Vedurmudi, J.~Goulet, J.~Christensen-Dalsgaard, B.~A. Young, R.~Williams,
  and J.~L. van Hemmen.
\newblock How internally coupled ears generate temporal and amplitude cues for
  sound localization.
\newblock \emph{Phys. Rev. Lett.}, 116\penalty0 (2):\penalty0 028101,
  2016{\natexlab{a}}.

\bibitem[Vedurmudi et~al.(2016{\natexlab{b}})Vedurmudi, Young, and van
  Hemmen]{ICE3}
A.~P. Vedurmudi, B.~A. Young, and J.~L. van Hemmen.
\newblock Internally coupled ears: mathematical structures and mechanisms
  underlying {ICE}.
\newblock \emph{Biol. Cybern.}, 110\penalty0 (4--5):\penalty0 359--382,
  2016{\natexlab{b}}.

\bibitem[Vossen et~al.(2010)Vossen, {Christensen-Dalsgaard}, and van
  Hemmen]{ICE1}
C.~Vossen, J.~{Christensen-Dalsgaard}, and J.~L. van Hemmen.
\newblock An analytical model for internally coupled ears.
\newblock \emph{J. Acoust. Soc. Am.}, 128\penalty0 (2):\penalty0 909--918,
  2010.

\bibitem[Williams(1999)]{Williams}
E.~Williams.
\newblock \emph{Fourier Acoustics: Sound Radiation and Nearfield Acoustic
  Holography}.
\newblock Academic, London, 1999.

\bibitem[Yosida(1980)]{Yosida}
K.~Yosida.
\newblock \emph{Functional Analysis}.
\newblock Springer, Berlin, 6th edition, 1980.

\bibitem[Zeidler(1999)]{Zeidler2}
E.~Zeidler.
\newblock \emph{Applied Functional Analysis: Applications to Mathematical
  Physics}.
\newblock Springer, Berlin, 1999.

\bibitem[Zhang et~al.(2006)Zhang, Hallam, and Christensen-Dalsgaard]{Chris3}
L.~Zhang, J.~Hallam, and J.~Christensen-Dalsgaard.
\newblock Modelling the peripheral auditory system of lizards.
\newblock In {S. Nolfi et al.}, editor, \emph{From Animals to Animats 9}, pages
  65--76, Berlin, 2006. Springer.

\end{thebibliography}
\end{document}